\documentclass[journal]{IEEEtran}

\usepackage{graphicx}
\usepackage{stackengine}

\ifCLASSINFOpdf
\else
\fi
\usepackage{amsmath}
\usepackage{amsfonts} 
\usepackage{hyperref}
\usepackage{subfigure}
\usepackage{algorithm}
\usepackage{algpseudocode}
\ifCLASSOPTIONcompsoc
\else
\fi
\usepackage{caption}
\usepackage[labelformat=simple]{subcaption}
 
\usepackage{dsfont}
\usepackage[table,xcdraw]{xcolor}
\usepackage{textcase}
\makeatletter

\renewcommand{\section}{\@startsection{section}{1}{\z@}{-3.5ex \@plus -1ex \@minus -.2ex}{2.3ex \@plus.2ex}{\normalfont\normalsize\centering\bfseries\MakeUppercase}}
\renewcommand{\subsection}{\@startsection{subsection}{2}{\z@}{-3.25ex\@plus -1ex \@minus -.2ex}{1.5ex \@plus .2ex}{\normalfont\normalsize\MakeUppercase}}
\makeatother
\usepackage{paralist}

\begin{document}

\title{Super-LoRa: Enhancing LoRa Throughput via Payload Superposition}
\author{Salah~Abdeljabar,~\IEEEmembership{Graduate~Student~Member,~IEEE,}
        and Mohamed-Slim~Alouini,~\IEEEmembership{Fellow,~IEEE}
\thanks{Salah Abdeljabar and Mohamed-Slim Alouini are with the Computer, Electrical and Mathematical Science and Engineering Division, King Abdullah University of Science and Technology (KAUST), Thuwal 23955-6900, Saudi Arabia (e-mail: salah.abdeljabar@kaust.edu.sa; slim.alouini@kaust.edu.sa).
 }
}

\markboth{IEEE Internet of Things Journal}%
{Shell \MakeLowercase{\textit{et al.}}: Bare Demo of IEEEtran.cls for IEEE Journals}

\maketitle

\begin{abstract}
This paper presents \textit{Super-LoRa}, a novel approach to enhancing the throughput of LoRa networks by leveraging the inherent robustness of LoRa modulation against interference. By superimposing multiple payload symbols, \textit{Super-LoRa} significantly increases the data rate while maintaining lower transmitter and receiver complexity. Our solution is evaluated through both simulations and real-world experiments, showing a potential throughput improvement of up to $5\times$ compared to standard LoRa. This advancement positions \textit{Super-LoRa} as a viable solution for data-intensive IoT applications such as smart cities and precision agriculture, which demand higher data transmission rates.
\end{abstract}

\begin{IEEEkeywords}
LoRa, Superposition, Software Defined Radio (SDR).
\end{IEEEkeywords}

\IEEEpeerreviewmaketitle
\section{Introduction}
\subsection{Motivation}
\par
The adoption of Internet of Things (IoT) technology has surged rapidly in recent years. According to \cite{frost2023}, approximately 41.76 billion IoT-connected devices were forecasted globally in 2023. Low-Power Wide-Area Network (LPWAN) technologies such as LoRa, Sigfox, and NB-IoT are recognized as the primary solutions for IoT connectivity, particularly for applications requiring long-range communication and low power consumption \cite{bccampus2023}. LoRa, in particular, is favoured for its ability to provide an extensive range with minimal power use, making it ideal for use cases such as smart cities \cite{andrade2019comprehensive, piechowiak2023lorawan} and smart agriculture \cite{semtech_lora_smart_agriculture, walter2017smart}. LoRa leverages Chirp Spread Spectrum (CSS) over narrow-band channels, offering data rates ranging from 0.146 kbps to 37.5 kbps depending on radio configuration \cite{askhedkar2023lora}. Therefore, LoRa is primarily adopted in low-data-rate applications due to these limitations.

\par
The growing demand for higher data transmission over IoT networks, including the need to transmit multimedia traffic (e.g., text, images), presents new challenges. For instance, smart agriculture applications may employ cameras to monitor crops, generating large bursts of traffic that require higher throughput \cite{vasisht2017farmbeats}. Similarly, environmental sensors monitoring critical phenomena (such as volcanic activity) demand rapid delivery of high-volume data, necessitating higher throughput communication over long distances.

\par
Despite LoRa’s advantages, including its long range and low power consumption, its relatively low data rates make it difficult to meet the needs of these emerging, data-intensive IoT applications. The limited throughput of LoRa also forces devices to remain active for longer durations, increasing energy consumption and reducing battery life \cite{phan2023}. \textcolor{black}{Moreover, the effective data rate in LoRa networks is further constrained by the low duty-cycle regulations of the industrial, scientific, and medical (ISM) band (e.g., $\leq 1\%$ in LoRaWAN \cite{LoRaWANDevelopers}). These limitations present significant challenges in supporting high-volume IoT traffic, such as the transmission of compressed images, audio, or large text files \cite{wei2020comparison, zhang2021lorawan, pham2016low} over LoRa networks.}

\par
On the same front, LoRa supports Adaptive Data Rate (ADR), a feature in LoRaWAN networks that dynamically adjusts the data rate based on channel conditions \cite{LoRaWANDevelopers}. The data rate in LoRa is mainly determined by two factors: the spreading factor (SF) and the modulation bandwidth (BW). Larger BW and smaller SF allow for higher data rates but reduce receiver sensitivity. Conversely, smaller BW and larger SF increase receiver sensitivity but lower the data rate, stretching transmission time. ADR aims to optimize this trade-off by adapting to the channel’s signal-to-noise ratio (SNR), with higher SNR allowing for faster data rates. However, in practice, the throughput gains achieved by ADR remain limited, particularly for IoT data-intensive applications \cite{xia2022hylink}.

\subsection{Contribution}

\par
The main contribution of this article is a novel LoRa transmission technique, called \textit{Super-LoRa}, which enables the superposition of multiple payload symbols to improve LoRa throughput. Our experiments indicate that \textit{Super-LoRa} can enhance throughput by up to $5\times$ compared to standard LoRa. This design leverages the robustness of LoRa modulation against interference, allowing a single LoRa node to transmit multiple superimposed symbols concurrently. As a result, collisions can be effectively decoded by standard receivers with minimal added complexity.

Our design draws inspiration from recent advances in LoRa collision resolution techniques, such as \textit{Choir} \cite{eletreby2017empowering}, \textit{mLoRa} \cite{wang2019mlora}, \textit{CIC} \cite{shahid2021concurrent}, \textit{CoLoRa} \cite{tong2020colora}, \textit{NScale} \cite{tong2020combating}, and \textit{Ftrack} \cite{xia2019ftrack}. These works primarily focus on exploiting LoRa’s modulation properties (such as time and frequency features) to enable LoRa gateways to resolve collisions and concurrently decode packets from multiple nodes, thereby improving the gateway throughput and the overall network capacity. However, these approaches often require greater receiver complexity, increased power consumption, and higher link budgets \cite{mishra2023openlora}, \textcolor{black}{let alone that the data rates per link are still limited.} 

\par
Our approach significantly increases throughput between communicating nodes while retaining LoRa’s lower transmitter and receiver complexity. We demonstrate that this is achievable by superimposing multiple payload symbols while retaining the same packet overhead (preamble, header, CRC, etc.) as a standard LoRa packet. Additionally, despite the received superimposed symbols at the receiver, standard LoRa demodulation can decode them with high reliability, requiring minimal, if any, additional complexity at the receiver. The contributions of this work can be summarized as follows:
\begin{itemize}
    \item We introduce \textit{Super-LoRa}, a novel transmission method that leverages the inherent robustness of LoRa modulation against interference by enabling the superposition of multiple payload symbols, significantly improving LoRa throughput.
    \item We examine the design considerations of \textit{Super-LoRa} and provide simulation results to validate its performance.
    \item We implemented \textit{Super-LoRa} on Software Defined Radios (SDR)  and evaluated its performance in both indoor and outdoor setups, demonstrating up to a $5\times$ throughput improvement compared to standard LoRa.
\end{itemize}

\section{Primer on LoRa communication}
LoRa is recognized as the most widely used LPWAN communication technique. According to market analysis, LoRa accounts for approximately 41\% of global LPWAN connections, which highlights its leading position in the LPWAN landscape \cite{iotanalytics_lpwan_market}. LoRa is also considered as a prominent choice for IoT applications due to its wide coverage and low power consumption, making it suitable for various deployments in smart cities and environmental monitoring systems  \cite{ayoub2023requirements}. In this section, we provide an overview of the LoRa physical layer, including the transmitted and received signals.

\subsection{LoRa Transmitter}
\begin{figure}[h!]
    \centering
    \includegraphics[width=0.9\linewidth]{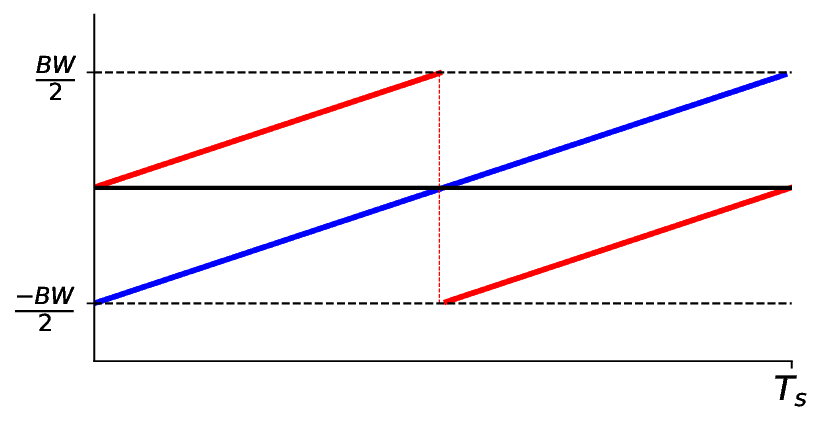}
    \caption{LoRa symbols for $SF = 8$ with different modulated symbols; $a = 0$ (plot in blue) and $a = 128$ (plot in red).}
    \label{fig:LoRa_Symbol}
\end{figure}

LoRa modulation employs a unique technique known as CSS, which is a form of spread spectrum technology. This method involves the transmission of data using chirp signals, which are frequency-modulated waveforms that increase or decrease in frequency over time. CSS spreads the signal across a wider bandwidth, which enhances the robustness of the communication against noise and interference \cite{de2020phase}. In CSS, symbols are modulated as up-chirps signals whose frequency increases linearly over a symbol duration $T_{s}$ and predefined bandwidth $BW$. Every symbol in LoRa is derived by cyclically shifting a base up-chirp $C(t)$ around $BW$ to modulate different data symbols. A mathematical representation of LoRa signal can be represented as follows:
\begin{align}
    S(t, f_{sym}) &= e^{j2\pi(\frac{{BW}^2}{2^{SF}}t - \frac{BW}{2} )t} \boldsymbol{\cdot} e^{j2\pi f_{sym}t} \nonumber \\
    S(t, f_{sym}) &= C(t) \boldsymbol{\cdot} e^{j2\pi f_{sym}t}, \quad 0 \leq t \leq T_{s}
    \label{equ:basic_LoRa}
\end{align}
where $SF$ corresponds to the \textit{Spreading Factor} of LoRa modulation, and the symbol time is $T_s = \frac{2^{SF}}{BW}$. $f_{sym}$ in \eqref{equ:basic_LoRa} is uniformly distributed over the signal $BW$ such as $f_{sym} = a\frac{BW}{2^{SF}}$  where $a \in \{ 0, 1, \dots, 2^{SF} - 1\}$, and $SF \in \{7, 8, 9, 10, 11, 12\}$ is fixed during the packet transmission and controls, together with $BW$, the transmission data rate. Figure \ref{fig:LoRa_Symbol} shows an example of LoRa symbols with different starting frequencies. Each symbol in LoRa can encode $2^{SF}$ bits so the symbol and bit rates for LoRa, $R_s$ and $R_b$, respectively, can be expressed as:
\begin{align}
    &R_s = \frac{1}{T_s} = \frac{BW}{2^{SF}} &\quad \text{symbols/s}\\
    &R_b = SF \boldsymbol{\cdot} R_s = SF \frac{BW}{2^{SF}} &\quad  \text{ bits/s}
\end{align}

\subsection{LoRa Receiver}
A LoRa receiver can demodulate an incoming chirp as follows. A LoRa receiver first multiplies the received signal by \textit{conjugate} of the base upchirp denoted by $C^*(t)$ over a window aligned with the symbol boundaries. This step de-chirps the signal and transforms it into a single tone. Then, it performs a Fast Fourier Transform (FFT) on the multiplied signal and searches for the largest power peak in FFT bins to demodulate a symbol. Mathematically, the demodulation of LoRa is expressed as:
\begin{align}
    C^*(t) \boldsymbol{\cdot} S(t, f_{sym}) = e^{j2\pi f_{sym}t}
    \label{equ:demodulated_LoRa}
\end{align}
Essentially, the FFT of the noiseless demodulated signal in \eqref{equ:demodulated_LoRa} produces one peak in the FFT bins that corresponds to the transmitted symbol $f_{sym}$.

\par
Since de-chirping requires the down-chirp to align with the received signal, LoRa utilizes a preamble to determine the boundaries of the received symbols. LoRa preamble comprises of a sequence of $N \in \{6, \dots, 65535 \}$ base up-chirps $C(t)$, followed by two \textit{SYNC} up-chirp symbols, and then $2.25$ down-chirps $C^*(t)$. LoRa receiver continuously de-chirps and performs FFT on the received signal until it detects the $N$ consecutive base up-chirps. Then, the \textit{SYNC} words, together with the down-chirps, help locate the packet’s boundary position.

\section{Super-LoRa Design}

\subsection{Overview}
\par
The idea behind \textit{Super-LoRa} builds on the observation that LoRa chirps exhibit a high level of immunity to interference from other LoRa chirps. This is also supported by the theoretical basis of LoRa modulation, where it has been demonstrated that standard LoRa demodulators can decode symbols even in the presence of interferers with the same SF \cite{afisiadis2019error}. 
The key question that \textit{Super-LoRa} is motivated by: what if the \textit{interference} is intentionally introduced by the transmitter to enable concurrent payload symbol transmission? If receivers can decode packets with overlapping symbols, we could intentionally superimpose payload data to increase throughput, which is crucial for IoT applications requiring higher data rates.
\par
\begin{figure}[h!]
    \centering
    \includegraphics[width=1.0\linewidth]{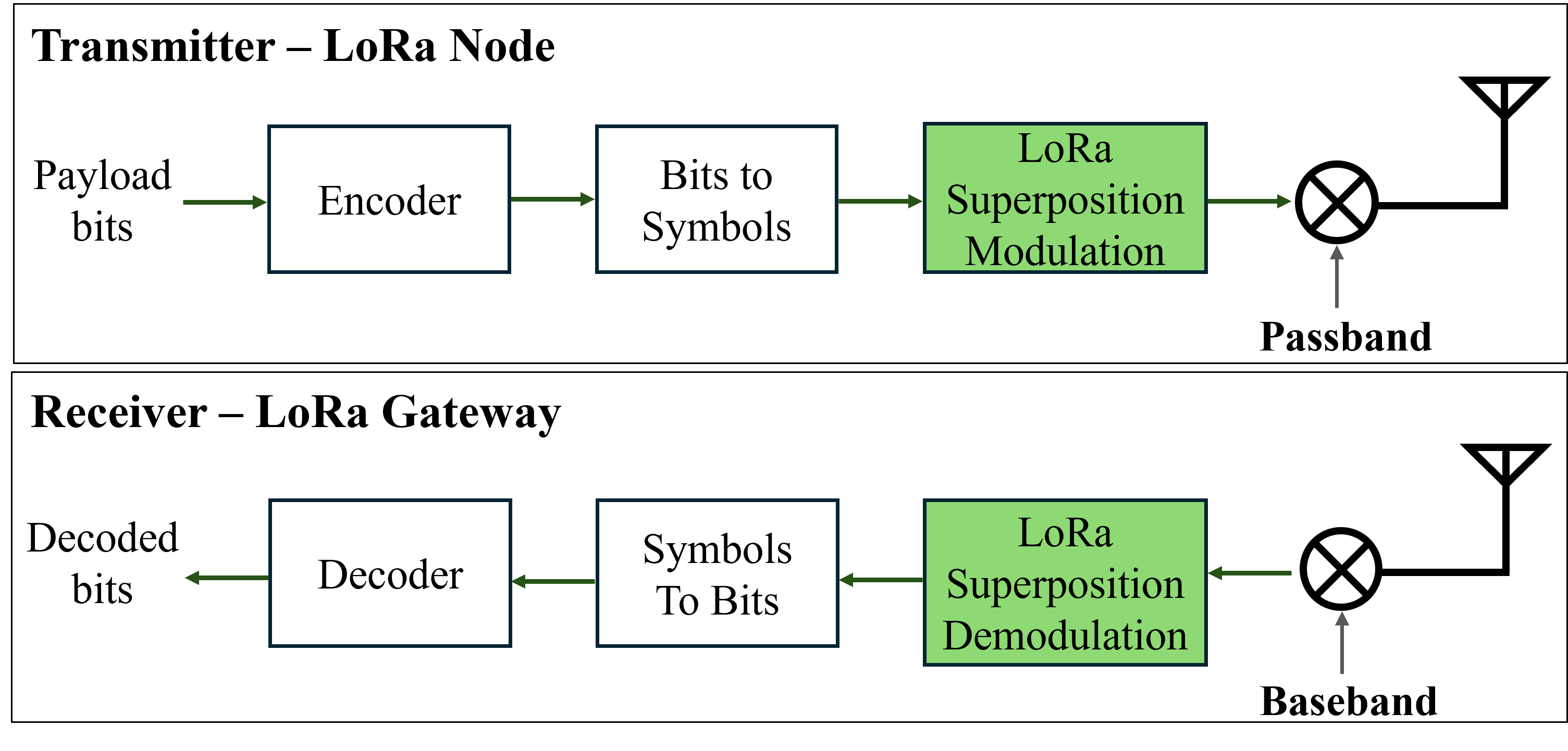}
    \caption{Illustration of \textit{Super-LoRa} design. \textit{Super-LoRa} reuses standard LoRa encoder and decoder, and \textcolor{black}{modify the modulator and demodulator to superimpose multiple payload symbols together and decode them, respectively.}}
    \label{fig:LoRa_Design}
\end{figure}

We propose a technique called \textit{Super-LoRa}, which leverages LoRa chirps’ inherent interference resilience to superimpose $K$ payload symbols. \textit{Super-LoRa} aims to balance increased data rates with maintaining low complexity at both the transmitter and receiver. As illustrated in Fig. \ref{fig:LoRa_Design}, \textit{Super-LoRa} design utilizes a standard LoRa encoder and decoder while modifying the design of the modulator and demodulator \textcolor{black}{to implement the superposition modulation and demodulation, respectively}. Based on channel conditions (e.g., SNR), the modulator decides how many symbols to embed within the chirp duration, superimposing part of the symbols before transmission. 
\par
\begin{figure}[h!]
    \centering
    \includegraphics[width=1.0\linewidth]{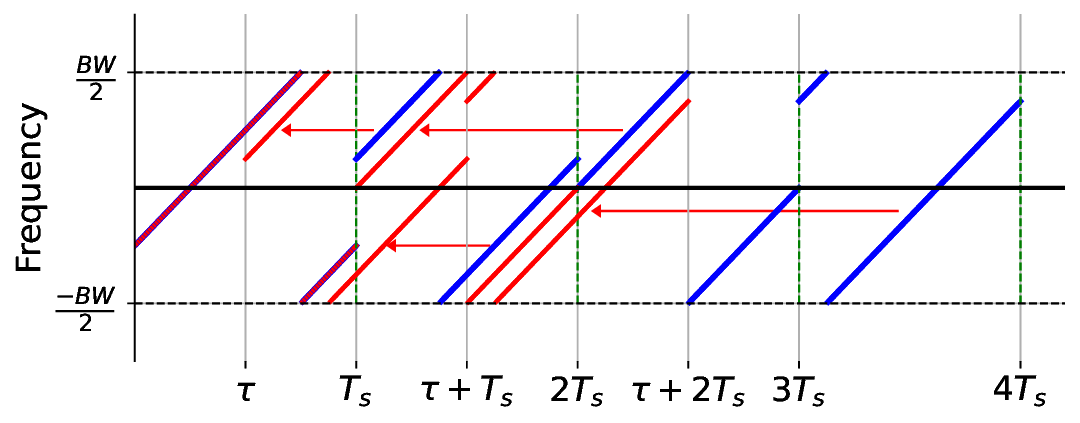}
    \caption{Illustration of LoRa superimposed payload symbols with superposition order $K = 2$ and $\tau = \frac{T_s}{2}$. \textcolor{black}{In blue, four consecutive symbols transmitted over $4T_s$ transmission time. In red, \textit{Super-LoRa} transmits the same four symbols over $2T_s + \tau$ transmission period ($\frac{T_s}{2}$ overlap between consecutive symbols), showing great potential of throughput improvement.}}
    \label{fig:LoRaSuperposition}
\end{figure}

Before transmission, \textit{Super-LoRa} appends the standard LoRa packet header (e.g., preamble, Start Frame Delimiter (SFD)), ensuring compatibility with existing receivers. At the receiver side, \textit{Super-LoRa} uses the standard LoRa technique to detect the packet’s start and demodulates the signal using FFT peak detection without additional interference mitigation steps. \textcolor{black}{To allow the receiver to decode the superimposed payload symbols, \textit{Super-LoRa} introduces a superposition time delay, denoted as $\tau$, which controls the level of payload symbol superposition and acts as a signature for the receiver to decode the superimposed symbols.
For example, in Fig. \ref{fig:LoRaSuperposition}, four payload symbols are transmitted by shifting the chirps closer in time by a factor $\tau$. In this case, the total transmission time was reduced from $4T_s$ to $2T_s + \tau$, showing the potential of improving the throughput over the packet transmission time.} 
\par
In summary, \textit{Super-LoRa} pushes the limits of LoRa’s throughput by increasing data rates up to $K$ times without introducing additional packet overhead or channel access delays. Importantly, \textit{Super-LoRa} remains backwards-compatible with legacy LoRa devices, ensuring seamless communication across networks.

\subsection{Super-LoRa Modulation}
\par
\textit{Super-LoRa} superimposes $K$ payload symbols to transmit them concurrently over the channel. However, when payload symbols are directly superimposed, a standard LoRa receiver cannot distinguish between them. To address this, our design introduces a \textit{time-delay} signature that allows the receiver to decode the superimposed symbols. This is based on the observation that standard LoRa decoders are highly resilient to \textit{non-aligned} interference (i,e, symbols with different start times). \textit{Super-LoRa} leverages this by adding a \textit{time-delay} between superimposed payload symbols, effectively using the delay as a signature for decoding. Mathematically, the \textit{Super-LoRa} modulated signal for $K$ payload symbols can be represented as follows:
\begin{align}
    x(t) &= \sum_{i = 0}^{K-1}{\sqrt{P_i} \boldsymbol{\cdot} S(t-\tau \boldsymbol{\cdot} i, f_{sym}(i))} \boldsymbol{\cdot} W(\frac{t-\tau \boldsymbol{\cdot} i}{T_s})  \label{equ:SuperPosition_Modulation} \\
    W(t) &= 
    \begin{cases} 
    1, & \text{if } 0 \leq t \leq 1, \\
    0, & \text{otherwise}.
    \end{cases} \\
    & 0 < K \boldsymbol{\cdot} \tau < T_s, \label{equ:delay_condition}
\end{align}
where $S(t-\tau \boldsymbol{\cdot} i, f_{sym}(i))$ and $P_i$ represent the symbol and power of the $i^{th}$ payload symbols, respectively. Here, $\tau$ denotes the time delay between each modulated payload symbols. Importantly, $\tau$ is a function of the superposition order $K$, which must be agreed upon by both the transmitter and receiver prior to transmission. For simplicity, we assume $\tau$ to be constant for all transmitted payloads, although it can be tuned based on the required modulation complexity. As to be discussed in Sections \ref{Power_Allocation} and \ref{Time_Delay}, both $\tau$ and $P_i$ are selected to optimize the reliability of the modulated LoRa payload symbols.
\par
\begin{figure}[h!]
    \centering
    \includegraphics[width=1.0\linewidth]{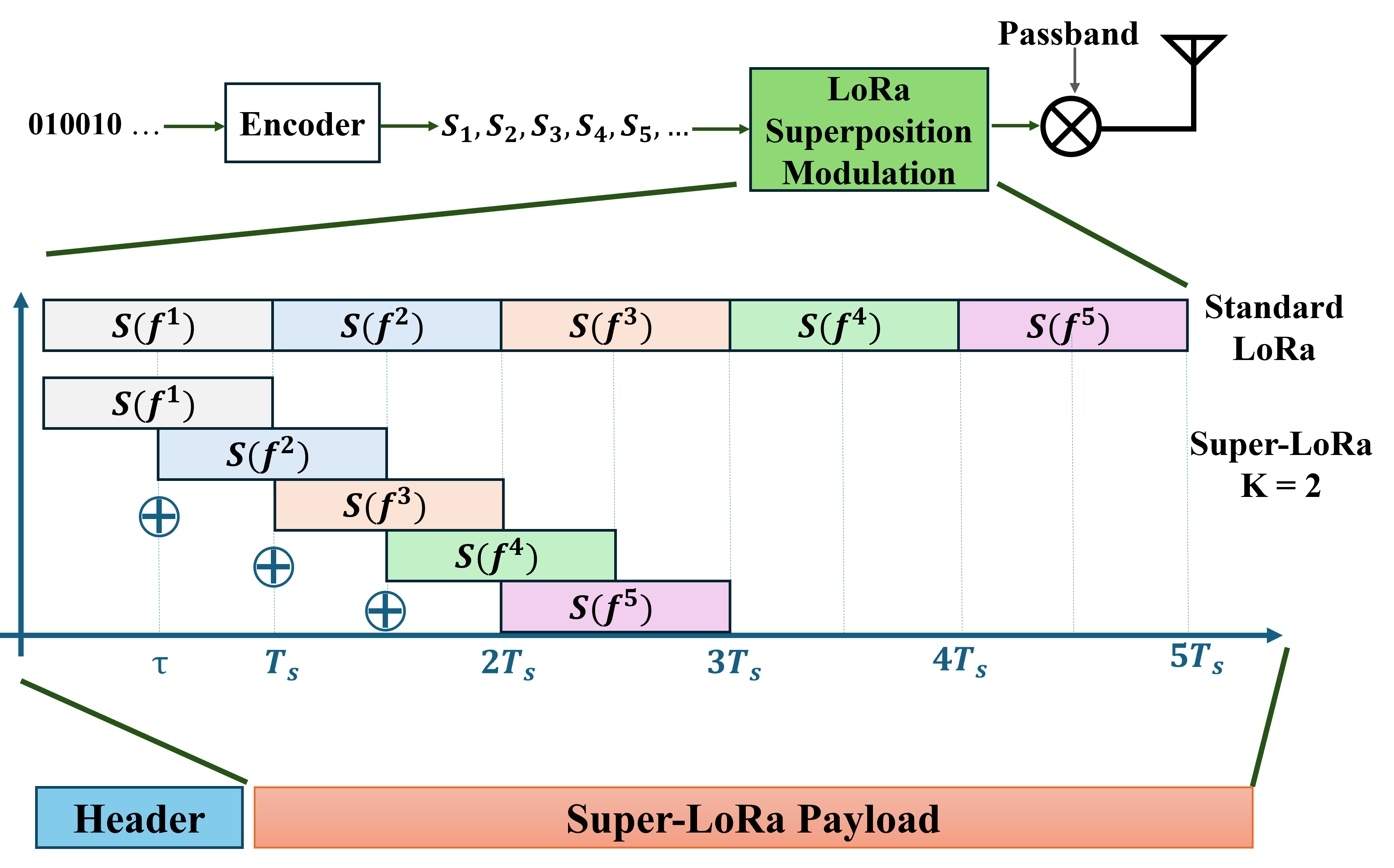}
    \caption{Illustration of \textit{Super-LoRa} payload superposition compared to standard LoRa. \textit{Super-LoRa} superimposes several payload symbols without introducing an extra overhead on the packet structure (utilize the same header as standard LoRa). \textcolor{black}{In this example, \textit{Super-LoRa} utilizes $K = 2$ with $\tau = \frac{T_s}{K}$.} }
    \label{fig:LoRa_Design_Modulator}
\end{figure}

The condition in Equation \eqref{equ:delay_condition} ensures that all superimposed chirps begin within the chirp duration $T_s$, maximizing the potential for superposition. Notably, when $K=1$, Equation \eqref{equ:SuperPosition_Modulation} reverts to the standard LoRa chirp, transmitting only one payload symbol within $T_s$. For $K>1$, each window of chirp duration contains one aligned payload and $2 \boldsymbol{\cdot}(K-1)$ non-aligned interfering payloads located at multiples of $\tau$. 
Additionally, it’s important to note that the superimposed symbol in Equation \eqref{equ:SuperPosition_Modulation}  extends beyond the chirp time $T_s$ by $K \boldsymbol{\cdot} \tau$, effectively creating interference for the subsequent chirp symbols. The details of the implementation of the modulator are discussed in Section \ref{Design_Consideration}. Figure \ref{fig:LoRa_Design_Modulator} shows \textcolor{black}{a comparison between a payload of standard LoRa and \textit{Super-LoRa} with $K = 2$ ($\tau = \frac{T_s}{K}$)}. It is shown in Fig. \ref{fig:LoRa_Design_Modulator} that \textit{Super-LoRa} superimpose several payload symbols without introducing an extra overhead on the packet structure (i,e, keep the same header as the one for standard LoRa). \textcolor{black}{This ensures backward compatibility of \textit{Super-Lora} with standard LoRa, and maintaining the detection and synchronization capabilities of standard LoRa through the packet header.}  

\subsection{Super-LoRa Demodulation} \label{sec:super-lora-demodulation}
\par
Following the payload symbols superposition presented in Equation \eqref{equ:SuperPosition_Modulation}, the received signal during each chirp duration $T_s$ will contain interference contributions from $K-1$ payload symbols to be transmitted \textit{next}, as well as $K-1$ superimposed payload symbols \textit{previously} transmitted. Unlike typical collision resolution algorithms, the number and relative time shifts of these superimposed payload symbols are predefined and agreed upon by both the transmitter and receiver, which are determined by $K$ and $\tau$.
\par
\begin{figure}[h!]
    \centering
    \includegraphics[width=1.0\linewidth]{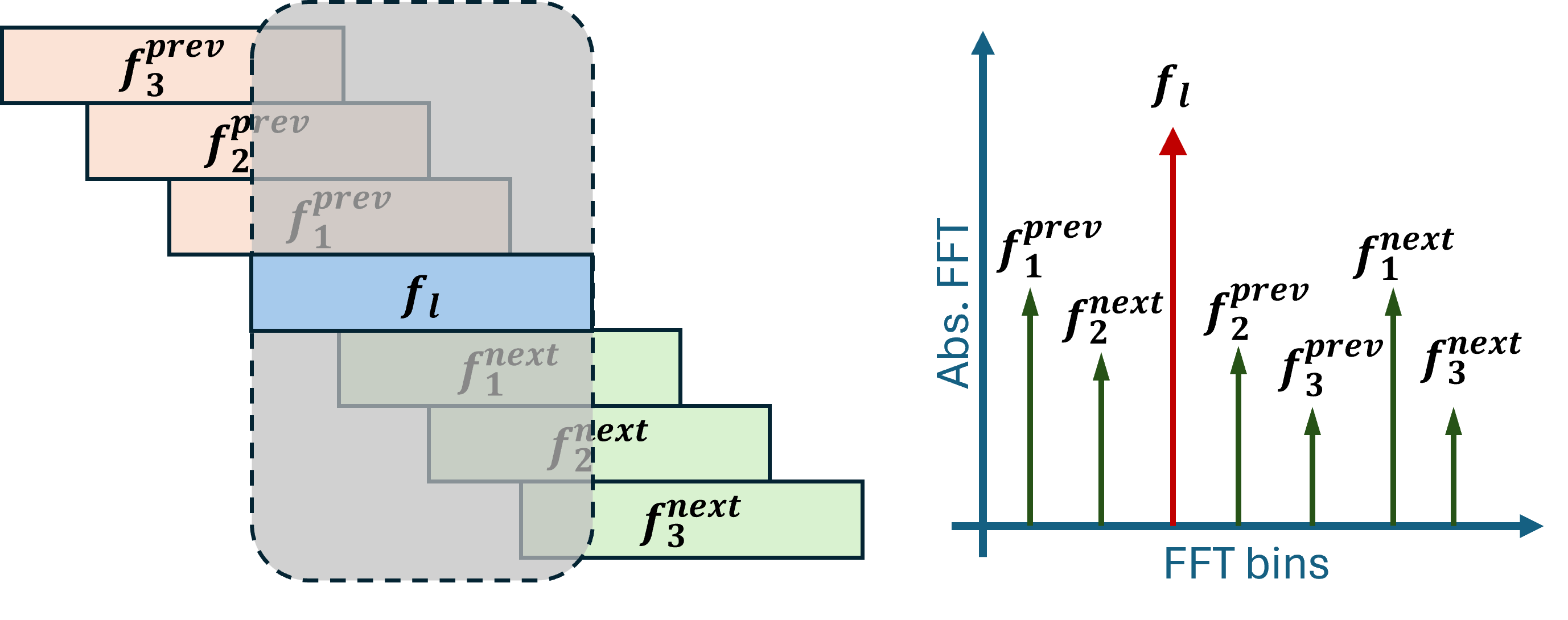}
    \caption{Once the receiver aligns the decoding window with the first payload symbol, all subsequent chirps will follow sequentially, with sliding the decoding window at a multiple of time delay $\tau$. The $l^{th}$ payload symbol exhibits interference from $K-1$ payloads to be decoded next, as well as $K-1$ decoded earlier. In this example, the superposition order is $K=4$, with $2\times(4-1)=6$ interfered peaks seen on the spectrum. The peaks of the interfered symbols are proportional to the time overlap with the target symbol.}
    \label{fig:LoRa_Design_Demodulator}
\end{figure}

We assume that the receiver has accurately determined the symbol boundaries of the first payload chirp using standard preamble detection techniques (all subsequent symbol chirps will follow sequentially, with a time delay $\tau$). Additionally, it is assumed that Carrier Frequency Offset (CFO) and Sample Timing Offset (STO) have been compensated for, following methods presented in the literature \cite{bernier2020low, shahid2021concurrent, xia2023pcube}.  Without loss of generality, the received signal $r(t)$ of the $l^{th}$ payload symbol is given by:
\begin{align}
    &r(t) = \alpha_l S\left(t, f_{sym}(l)\right) + I_{next}(t) + I_{prev}(t) + n(t)  \label{equ:Received_Signal}\\
     &I_{prev}(t) =\sum_{i = 1}^{K-1}{\alpha_i^{prev} \boldsymbol{\cdot} S\left(t + T_s -\tau \boldsymbol{\cdot} i, f^{prev}_{sym}(i)\right)} \boldsymbol{\cdot} W\left(\frac{t}{\tau \boldsymbol{\cdot} i}\right) \label{equ:Interference_Prev}\\ 
     &I_{next}(t) =\sum_{i = 1}^{K-1}{\alpha_i^{next} \boldsymbol{\cdot} S\left(t-\tau \boldsymbol{\cdot} i, f^{next}_{sym}(i)\right)} \boldsymbol{\cdot} W\left(\frac{t-\tau \boldsymbol{\cdot} i}{T_s-\tau \boldsymbol{\cdot} i}\right) \label{equ:Interference_Next}
\end{align}
where $\alpha_l, \alpha_i^{prev}$ and $\alpha_i^{next}$ are the amplitude of the received signal from for the $l^{th}$, $prev$, and $next$ payload symbols, respectively. The term $n(t)$ represents the AWGN noise, where $I_{prev}(t)$ and $I_{next}(t)$ represent the interference contribution from the $prev$, and $next$ payload symbols, respectively. The window function $W(t)$ captures the overlap region between the target symbol and the $prev$, and $next$ symbols, with widths proportional to $\tau$. Notably, the first $K-1$ payload symbols experience less prior interference, but for generality, we consider the case depicted in Equation \eqref{equ:Interference_Prev}.
The received signal in Equ. \eqref{equ:Received_Signal} is then passed through standard LoRa demodulation (de-chirping then FFT). The result produces $2\left(K-1\right)$ peaks, corresponding to the overlapping superimposed symbols, along with one peak for the target $l^{th}$ symbol. The equivalent Fourier transform of the received signal after de-chirping is given by:
\begin{align}
    &R(f) = \alpha_l \cdot sinc\left[T_s \left(f - f_{sym}\left(l\right)\right)\right] +\nonumber \\
    & \qquad \qquad \qquad \qquad \qquad I_{next}(f) + I_{prev}(f) + N(f) \label{equ:Received_Signal_Freq}\\
    &I_{prev}(f) =\sum_{i = 1}^{K-1}{\frac{\alpha_i^{prev}}{\tau \cdot i} \boldsymbol{\cdot} sinc\left[ \left(\tau \cdot i\right) \left(f - f^{prev}_{sym}(i) - \Delta f_i\right)\right]} \label{equ:Interference_Prev_Freq}\\ 
    &I_{next}(f) =\sum_{i = 1}^{K-1}{\frac{\alpha_i^{next}}{T_s - \tau \cdot i} \boldsymbol{\cdot}} \nonumber \\
    &\qquad \qquad \qquad sinc\left[ \left(T_s - \tau \cdot i\right) \left(f - f^{next}_{sym}(i) - \Delta f_i\right)\right] \label{equ:Interference_Next_Freq} \\
    &\Delta f_i = \left(\tau \cdot i\right) \cdot \left(\frac{BW}{2^{SF}}\right)
\end{align}
Note from Equs. \eqref{equ:Interference_Prev_Freq} and \eqref{equ:Interference_Next_Freq} that the peaks of the superimposed symbols $\frac{\alpha_i^{prev}}{\tau \cdot i}$ and $\frac{\alpha_i^{next}}{T_s - \tau \cdot i}$ depend on the received amplitudes (power) and the time overlap with the target symbol. In fact, the peak height is inversely proportional to the time overlap with the target symbol. To illustrate the idea more, we depict in Fig. \ref{fig:LoRa_Design_Demodulator} the received signal as described in Equ. \eqref{equ:Received_Signal_Freq}. It is clear that the peaks of the interfered symbols are proportional to the time overlap with the target symbol.
\par
\begin{figure}[h!]
    \centering
    \includegraphics[width=1.0\linewidth]{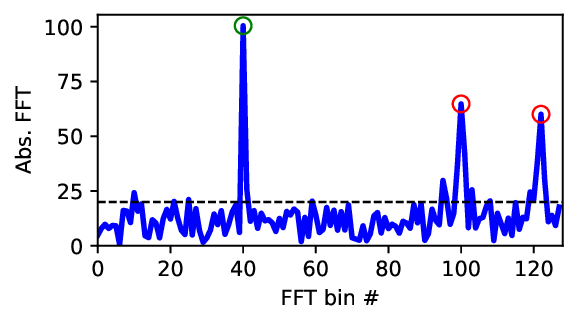}
    \caption{De-chirped superimposed LoRa signal with $K = 2$. The target signal has a peak at bin index $40$ (circled with green), and $SNR$ is $0$ dB. The transmitted signal has $SF=7, BW=125 \text{ KHz}$. The superpositon time delay $\tau$ was set to $\tau = \frac{Ts}{2}$ (this corresponds to $64$ samples for $SF=7$). The previous superimposed signal has a peak at bin $122$ while the next superimposed signal peak at bin $100$ (both circled with red).}
    \label{fig:InterferencePlot}
\end{figure}

Furthermore, in Fig. \ref{fig:InterferencePlot}, we depict an example of a demodulated superimposed signal with $K = 2$ payload symbols passing through AWGN channel with SNR $= 0$ dB (SNR is calculated as the ratio of the power of the superimposed signal to the noise power). The superposition time delay $\tau$ was set to $\tau = \frac{Ts}{2}$, with all payload symbols transmitted at equal power. It is evident that the target symbol has the highest peak, while the superimposed symbols exhibit lower peaks due to the overlap reduction. Although superimposed chirps can negatively affect the target chirp, the superposition time delay ensures that their peaks are scaled by the overlap time, resulting in lower peaks compared to the aligned target symbol. This demonstrates the potential of \textit{Super-LoRa} to carry multiple payload symbols concurrently, thereby significantly increasing data rates.
\par
\begin{figure}[h!]
    \centering
    \includegraphics[width=1.0\linewidth]{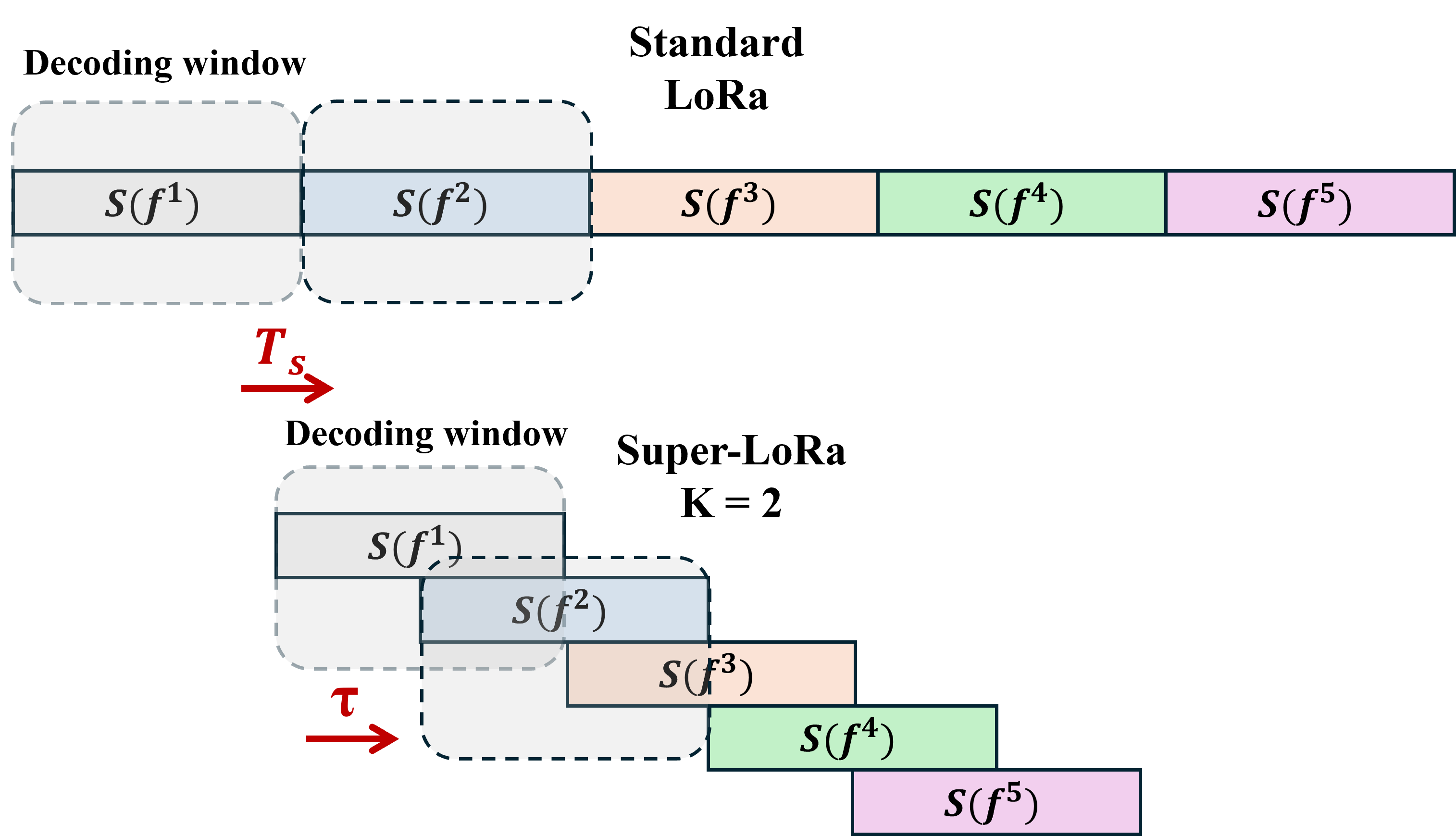}
    \caption{\textcolor{black}{ Illustration of \textit{Super-LoRa} decoding window compared to standard LoRa. The receiver aligns the decoding window with the first payload symbol, which is detected through the preamble as performed in standard LoRa. Instead of decoding each payload symbol every chirp time $T_s$, the receiver in \textit{Super-LoRa} decodes every $\tau$ seconds. This maintains the same processing capabilities while offering higher resource utilization.}}
    \label{fig:Decoding_Window}
\end{figure}

\textcolor{black}{The receiver aligns the decoding window with the first payload symbol, which is detected through the preamble as performed in standard LoRa. Next, instead of sliding the decoding window for each payload symbol every chirp time $T_s$, the receiver in \textit{Super-LoRa} decodes every $\tau$ seconds. This maintains the same processing capabilities while offering higher resource utilization. An illustration of how the decoding window moves between payload symbols in \textit{Super-LoRa} with $K=2$ compared to standard LoRa is shown in Fig. \ref{fig:Decoding_Window}.}
\subsection{Design Considerations and Limitations} \label{Design_Consideration}
\par 
While the previous section provided a concise overview of the changes introduced by \textit{Super-LoRa} to the modulator and demodulator compared to standard LoRa, several key aspects of our design remain to be elaborated. First, given a limited transmission power budget, what is the optimal power allocation among the superimposed payload symbols to maximize transmission reliability? 
\textcolor{black}{Next, how should these payload symbols be superimposed, i.e., how to choose the superposition time delay $\tau$ (or $K$)? Then, what are the performance Trade-offs presented in our design? Next, how \textit{Super-LoRa} selects the superposition order $K$?} Then, how should \textit{Super-LoRa}’s modulator be implemented to handle symbol superposition? Finally, how should the demodulator be realized in this proposed framework? In this section, we address these key questions.

\subsubsection{\textbf{Power Allocation}}\label{Power_Allocation}
\par
\textcolor{black}{
In \textit{Super-LoRa}, the superimposed payload symbols share the same total transmitted power, denoted as \( P_T \) (see Equ. \eqref{equ:SuperPosition_Modulation}). A key question we address in this part is how to allocate the power among these symbols to maximize communication reliability. According to our framework, the power budget is distributed among \( K \) symbols, satisfying the constraint:
\begin{equation}
    P_1 + P_2 + \dots + P_K = P_T.
\end{equation}
The power allocated to each symbol directly affects communication reliability, influencing how well peaks can be distinguished from one symbol to another. As seen in Eqs. \eqref{equ:Received_Signal_Freq}, \eqref{equ:Interference_Prev_Freq}, and \eqref{equ:Interference_Next_Freq}, the amplitude of each superimposed symbol determines whether it will be correctly decoded.
In each decoding window, \textit{Super-LoRa} symbols experience interference from other superimposed symbols, impacting the signal-to-interference ratio (SIR). 
Since the Symbol Error Rate (SER) is directly influenced by the SIR of each received symbol \cite{afisiadis2019error}, \textit{Super-LoRa} allocates power to maximize the minimum SIR across all superimposed payload symbols. Based on our design, the SIR for the \( l^{\text{th}} \) symbol decoding window (see Fig. \ref{fig:LoRa_Design_Demodulator}) is defined as:
\begin{equation} \label{Equ:SIR}
    \text{SIR}_l = \frac{P_l}{\sum_{\substack{i=1 \\ i \neq l}}^{K} P_i}.
\end{equation} 
Equation \ref{Equ:SIR} indicates that allocating more power to one symbol at the expense of others decreases the average SIR, making uniform allocation the optimal choice.
Therefore, given a total transmission power $P_T$, \textit{Super-LoRa} uses uniform power distribution across all symbols, i,e, $P_1=P_2=\dots=P_K=\frac{P_T}{K}$, to ensure the best SER performance.  
Appendix \ref{sec:Optimal_Power_Allocation_Proof} formally proves that uniform power allocation maximizes the minimum SIR across all superimposed payload symbols.
It is worth noting that the impact of the SIR on the communication reliability is to be thoroughly investigated in Section \ref{sec:performance_tradeoffs}, which has a similar analysis as the ones in \cite{elshabrawy2018analysis}.}
\subsubsection{\textbf{Superposition Time Delay}}\label{Time_Delay}
\par
\textcolor{black}{
A key aspect of \textit{Super-LoRa} is selecting the superposition time delay $\tau$, which determines the number of symbols superimposed within a single symbol duration $T_s$.
Like other LoRa parameters (SF, BW, etc.), both the transmitter and receiver must agree on $\tau$ before transmission. In our framework, we define $\tau$ as a fraction of the symbol time $T_s$, $\tau = \frac{T_s}{K}$, where $K$, called superposition order, is a number greater than or equal to 1 ($K \geq 1$) and determines the level of superposition selected in the design. 
Note that the selection of one of the parameters ($K$ or $\tau$) directly leads to the other, so we focus here on the selection of the superposition order, $K$.
When $K = 1$, there is no payload superposition and the transmission reverts to standard LoRa. $\tau$ formally indicates the duration where consecutive payload symbols would be superimposed (or overlapped). This overlap would create an interference with the adjacent symbols, which could potentially affect the communication reliability. Notably, increasing $K$ (or equivalently reducing $\tau$) increases symbol superposition, enhancing throughput but that comes at the expense of reduced communication reliability.
On the other hand, lower values of $K$ reduce the level of superposition, resulting in lower throughput gains but improved reliability. This is clear from looking at the received \textit{Super-LoRa} symbol frequency components shown in Equs. \eqref{equ:Received_Signal_Freq}, \eqref{equ:Interference_Prev_Freq}, and \eqref{equ:Interference_Next_Freq}. 
To simplify the design and parameter negotiation between the transmitter and receiver, we limit the selection of the superposition order $K$ to integers. This discrete formulation ensures that once the transmitter and receiver agree on the value of $K$, the superposition time delay $\tau$ is automatically determined. 
We can see that $\tau$, along with the allocated power, affects the FFT peak of each superimposed payload symbol. This will have a direct impact on the SER of the system and hence the receiver reliability. Therefore, $K$ plays a key role in the trade-off between improved throughput and compromised reliability. The details of this trade-off is presented in Section  \ref{sec:performance_tradeoffs}.}

\subsubsection{\textcolor{black}{\textbf{\textit{Performance Trade-offs}}}}\label{sec:performance_tradeoffs}
\par
\textcolor{black}{
\textit{Super-LoRa} leverages the inherent robustness of LoRa chirps against non-aligned interference to enhance throughput through controlled symbol superposition. This is achieved by overlapping consecutive symbols (see Fig. \ref{fig:LoRa_Design_Modulator} and Fig. \ref{fig:LoRa_Design_Demodulator}), which reduces packet transmission time while maintaining compatibility with standard LoRa receivers. At the receiver, decoding is performed by shifting the decoding window in steps of $\tau$, rather than the full symbol duration $T_s$ (see Fig. \ref{fig:Decoding_Window}). This approach requires minimal modifications to the LoRa receiver, preserving its low complexity and power efficiency.
} 

\par
\textcolor{black}{
A key trade-off in \textit{Super-LoRa} arises between throughput gains and communication reliability, governed by the parameter $K$. This trade-off is analyzed in detail in the following. First, the bit rate gains in \textit{Super-LoRa} can be quantified based on the superposition order $K$. 
Let $R_b$ denote the bit rate of standard LoRa (equivalent to \textit{Super-LoRa} with $K=1$). For a packet containing $M$ payload symbols, the bit rate of \textit{Super-LoRa}, denoted as $R_b^{\text{super}}$, can be given by:
\begin{equation}
    R_b^{\text{super}} = R_b \cdot \frac{M}{\frac{M-1}{K} + 1}.
\end{equation}
where for large payload packets and moderate superposition order $K$, the bit rate can be approximated as:
\begin{equation}\label{Equ:bit_rate_approx}
    R_b^{\text{super}} \approx R_b \cdot K,
\end{equation}
The approximation in Equ. \ref{Equ:bit_rate_approx} holds for larger payloads and relatively smaller values of $K$. This demonstrates that the bit rate improvement in \textit{Super-LoRa} scales linearly with $K$, offering significant throughput gains compared to standard LoRa.
}
\par
\begin{figure}[h!]
    \centering
    \includegraphics[width=1.0\linewidth]{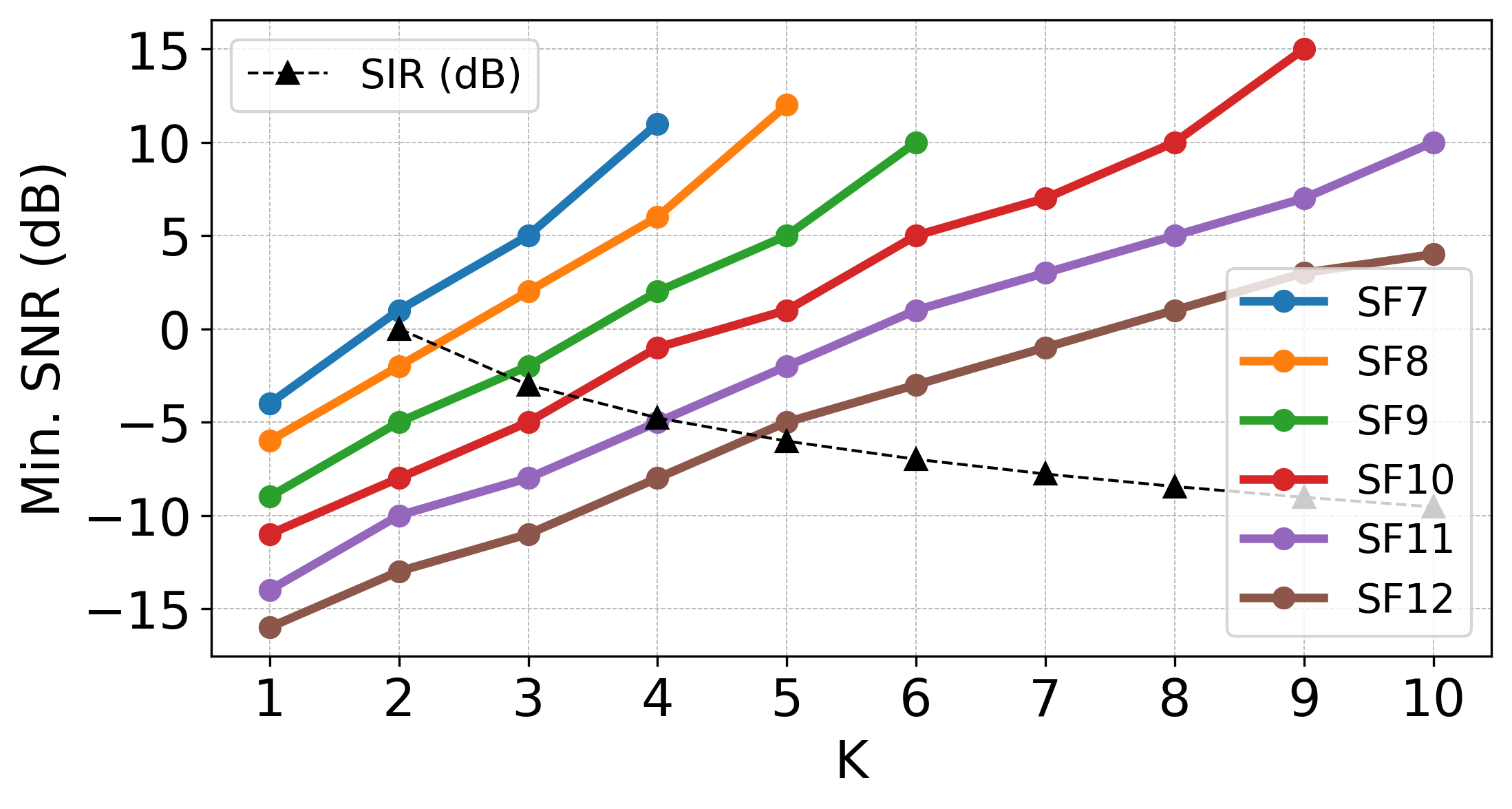}
    \caption{\textcolor{black}{ Minimum required SNR to achieve SER = $10^{-1}$ for each superposition order $K$. Note that there is a penalty of around 4-5 dB for each additional $K$ symbols superimposed. In black, we plot at the same figure the resulting SIR (defined as $SIR = \frac{1}{K-1}$) for each superposition order $K$.}}
    \label{fig:MinSNR_VS_K}
\end{figure}

\textcolor{black}{
However, the superposition of symbols introduces additional interference, which impacts communication reliability. As illustrated in Fig. \ref{fig:LoRa_Design_Demodulator}, each decoding window contains interference from $K-1$ previous and $K-1$ next payload symbols. This results in an effective interference equivalent to $K-1$ fully overlapped symbols, though, not time-aligned with the target symbol. Since \textit{Super-LoRa} employs standard LoRa demodulation without advanced interference cancellation techniques, this interference directly affects SIR and the SER performance at each decoding window \cite{afisiadis2019error}. The impact of such interference on LoRa’s demodulation performance has been extensively studied in prior works \cite{elshabrawy2018analysis, elshabrawy2018closed, afisiadis2019lora}. In our framework, since we adopt uniform power allocation among superimposed symbols, the SIR for each decoding window is given by:
\begin{equation} \label{equ:SIR_final}
    \text{SIR} = \frac{1}{K - 1}.
\end{equation}
The SIR, when expressed in decibels, exhibits a logarithmic decay with respect to $K$.
For example, when $K=2$, the SIR is $0$ dB, but for $K=3$, it drops to approximately $-3.01$ dB, and for $K=4$, it decreases further to $-4.77$ dB. This trend continues as $K$ increases, with the SIR reduction becoming less steep for larger values of $K$. We plot the SIR values based on Equ. \ref{equ:SIR_final} on Fig. \ref{fig:MinSNR_VS_K}. Importantly, this reduction in SIR as $K$ grows has a direct impact on the SER performance as highlighted next.}

\textcolor{black}{
To quantify the impact of interference on communication reliability, we follow a similar analysis as the one on \cite{elshabrawy2018analysis} and conduct a comprehensive simulation study of \textit{Super-LoRa} over a Rayleigh fading channel.  While our approach intentionally introduces controlled interference for throughput gains, the SER degradation follows similar patterns observed in prior studies that consider LoRa networks exhibiting same-SF interference \cite{afisiadis2019error, elshabrawy2018closed, afisiadis2019lora}.
Fig. \ref{fig:MinSNR_VS_K} presents the minimum required SNR to achieve a target SER performance of $10^{-1}$ for various values of $K$ and SF. This SER threshold is chosen based on practical considerations, as LoRa employs Forward Error Correction (FEC) and upper-layer coding schemes (e.g., Hamming codes) that can effectively correct errors below this level \cite{xia2022hylink}. 
Our results indicate that each additional superposition order $K$ increases the required SNR by approximately $4$–$5$ dB due to the additional interference. This aligns with prior findings \cite{afisiadis2019error}, which report similar SNR penalties for LoRa under interference conditions. Although the required SNR is smaller for higher SFs, the pattern remains consistent.
}
\par
\textcolor{black}{The increased SNR requirement imposes a limitation on the maximum communication range of \textit{Super-LoRa}. To address this, our design dynamically adjusts the superposition order $K$ based on channel conditions, ensuring an optimal balance between throughput and reliability. This adaptive mechanism is further investigated in Section \ref{sec:Super-LoRa-Order}.}

\textcolor{black}{
\subsubsection{\textbf{Superposition Order Selection}}\label{sec:Super-LoRa-Order}
\textit{Super-LoRa} dynamically selects the superposition order $K$ based on channel conditions to maximize throughput while maintaining reliable communication. The bit rate improvement in \textit{Super-LoRa} scales linearly with $K$, i.e., $R_b^{super} \approx R_b \cdot K$, where $R_b$ is the bit rate of standard LoRa. 
In Fig. \ref{fig:MinSNR_VS_K}, we analyze the minimum required SNR to achieve a target SER as a function of $K$. \textit{Super-LoRa} uses this knowledge to adaptively adjust $K$. For instance, in the case of LoRa with $SF7$, standard LoRa ($K=1$) requires an SNR of approximately $-4$ dB to achieve the target SER. However, if the SNR improves to around $0$ dB, \textit{Super-LoRa} can reliably operate with $K=2$, effectively doubling the bit rate. Generally, as observed in Fig. \ref{fig:MinSNR_VS_K}, although higher SFs require lower SNRs to support larger values of $K$, each additional superposition order introduces an approximate $5$ dB SNR penalty. 
\textit{Super-LoRa} exploits real-time link conditions to transmit at the highest possible bit rate by selecting the largest $K$ (denoted as $K_{\max}$) that satisfies the target SER performance. If the channel degrades, \textit{Super-LoRa} reduces $K$ accordingly, defaulting to $K=1$ (equivalent to standard LoRa) under poor conditions.
The adaptation mechanism operates as follows: Let $SNR_0^{SF}$ represent the minimum SNR required for standard LoRa ($K=1$) to achieve target SER at a given SF. For measured SNR values exceeding this baseline, based on our coarse estimations, the maximum viable superposition order $K_{\max}$ is determined by:
\begin{equation} \label{Equ:K_max}
    K_{\max} \approx \left\lfloor \frac{SNR - SNR_0^{SF}}{5} \right\rfloor + 1.
\end{equation}
Equation \eqref{Equ:K_max} ensures that \textit{Super-LoRa} dynamically adjusts $K$ based on the available SNR, maximizing data rates while maintaining robust communication.
It is worth noting that practical implementation requires coordination between transceivers to synchronize $K$ selection. While this could be achieved through mechanisms analogous to LoRaWAN’s ADR feature \cite{LoRaWANDevelopers}, protocol design constitutes a distinct research challenge beyond the scope of this paper. Our current focus establishes the fundamental performance boundaries and adaptation criteria, providing a foundation for subsequent system-level implementations.
}

\subsubsection{\textbf{Super-LoRa Modulator}}\label{Super-LoRa-Modulator}
\par
As illustrated in Fig. \ref{fig:LoRa_Design_Modulator}, \textit{Super-LoRa} introduces payload superposition into the modulator block while retaining the same encoder used in standard LoRa. First, \textit{Super-LoRa} sequentially accepts symbols for modulation and then applies standard LoRa modulation to convert these symbols into \textit{I-Q} samples of length $N_s$, where $N_s$ represents the number of samples corresponding to one chirp. However, instead of transmitting each set of $N_s$ samples immediately, \textit{Super-LoRa} buffers the modulated \textit{I-Q} samples based on the superposition order $K$. Once $K$ symbols have been modulated, superimposed, and buffered, the modulator sends the combined $N_s$ samples to the analog section for radio transmission.
Afterwards, more modulated samples are added to the end of the buffer while older samples are removed. The required buffer size depends on both the superposition order $K$ and the superposition delay $\tau$. Assuming $\tau = \frac{T_s}{K}$, which corresponds to $\lceil \frac{N_s}{K} \rceil$ samples, the buffer size should be $2\times N_s - \lceil \frac{N_s}{K} \rceil$. After transmitting every $N_s$ samples, the modulator discards the oldest $\lceil \frac{N_s}{K} \rceil$ samples and adds new modulated samples to the tail of the buffer. This process continues for each subsequent symbol until all payloads have been transmitted.
Importantly, the radio frequency (RF) chain of the transmitter remains unchanged, with no additional signal processing complexity introduced. This highlights the simplicity of the design, as no further modifications to standard LoRa modulators are required.

\subsubsection{\textbf{Super-LoRa Demodulator}}\label{Super-LoRa-Demodulator}
\par
Similar to the \textit{Super-LoRa} modulator, the demodulator uses buffering to process the received samples. Upon reception, the payload samples are buffered at the DSP stage. Each set of $N_s$ samples is then sent to a standard LoRa demodulator (de-chirping, followed by FFT). After decoding a symbol, the oldest $\lceil \frac{N_s}{K} \rceil$ samples are discarded, while new samples are added to the buffer. Unlike standard LoRa, where all $N_s$ samples are discarded after decoding, \textit{Super-LoRa} only removes part of them, keeping the rest for overlap. The buffer size required is $2\times N_s - \lceil \frac{N_s}{K} \rceil$ samples.

\section{Evaluation}
\par
In this part, we evaluate the efficacy of \textit{Super-LoRa} with various LoRa parameters. We first validate our framework using simulation over \textcolor{black}{Rayleigh fading channel, which is more representative of real-world wireless environments compared to AWGN channel}, before experimenting with real-world deployments. 
\textcolor{black}{In our study, we aimed to capture a wide range of SNR conditions, spanning from -15 dB to 20 dB, to reflect the practical and diverse channel conditions in which LoRa networks typically operate. LoRa is designed for power-constrained IoT devices with limited transmission power, and its long-range communication results in received SNR values that fall within the investigated range. Furthermore, our experimental validation, which included various channel environments (e.g., indoor and outdoor setups, line-of-sight (LoS) and non-line-of-sight (NLoS) conditions), yielded similar SNR values, supporting the relevance of our chosen range.} 

\subsection{Simulation}\label{sec:Simulations}
\par
In this part, we validate the proposed framework via simulation. We show the superiority of the proposed framework in terms of effective bitrate and highlight its main limitations. We simulated \textit{Super-LoRa} received signal presented in Equ. \eqref{equ:Received_Signal} over \textcolor{black}{ Rayleigh fading channel} with various superposition order $K$, SNR, and SF. \textcolor{black}{It is worth noting that we experimented \textit{Super-LoRa} with a typical AWGN channel; however, while considering Rayleigh fading introduces higher SER values compared to AWGN, the throughput improvement achieved by \textit{Super-LoRa} compared to standard LoRa remains consistent. This confirms the robustness of our framework in diverse channel environments.}

We assumed a perfect detection window for the target symbol. We record the SER, and the effective bitrate ($R_{\text{effective}}$) which was calculated as follows: 
\begin{align}
    R_{\text{effective}} & = R_{\text{gross}} \times \left(1- SER\right) & \text{ bits/s}\label{equ:effective_bitrate} \\
    R_{\text{gross}} &  \approx K \boldsymbol{\cdot}SF\boldsymbol{\cdot}\frac{BW}{2^{SF}} & \text{ bits/s}\label{equ:gross_bitrate}
\end{align}
Effective bitrate illustrates how an increase in SER directly impacts the system’s bitrate upper bound, emphasizing its importance in system design and performance evaluation. The results were averaged over $10,000$ independent realizations. All simulations consider LoRa transceivers to operate with $BW = 250$ KHz.

\par
\begin{figure}[h!]
	\centering
	\subfigure[]
	{\includegraphics[width=0.49\linewidth]{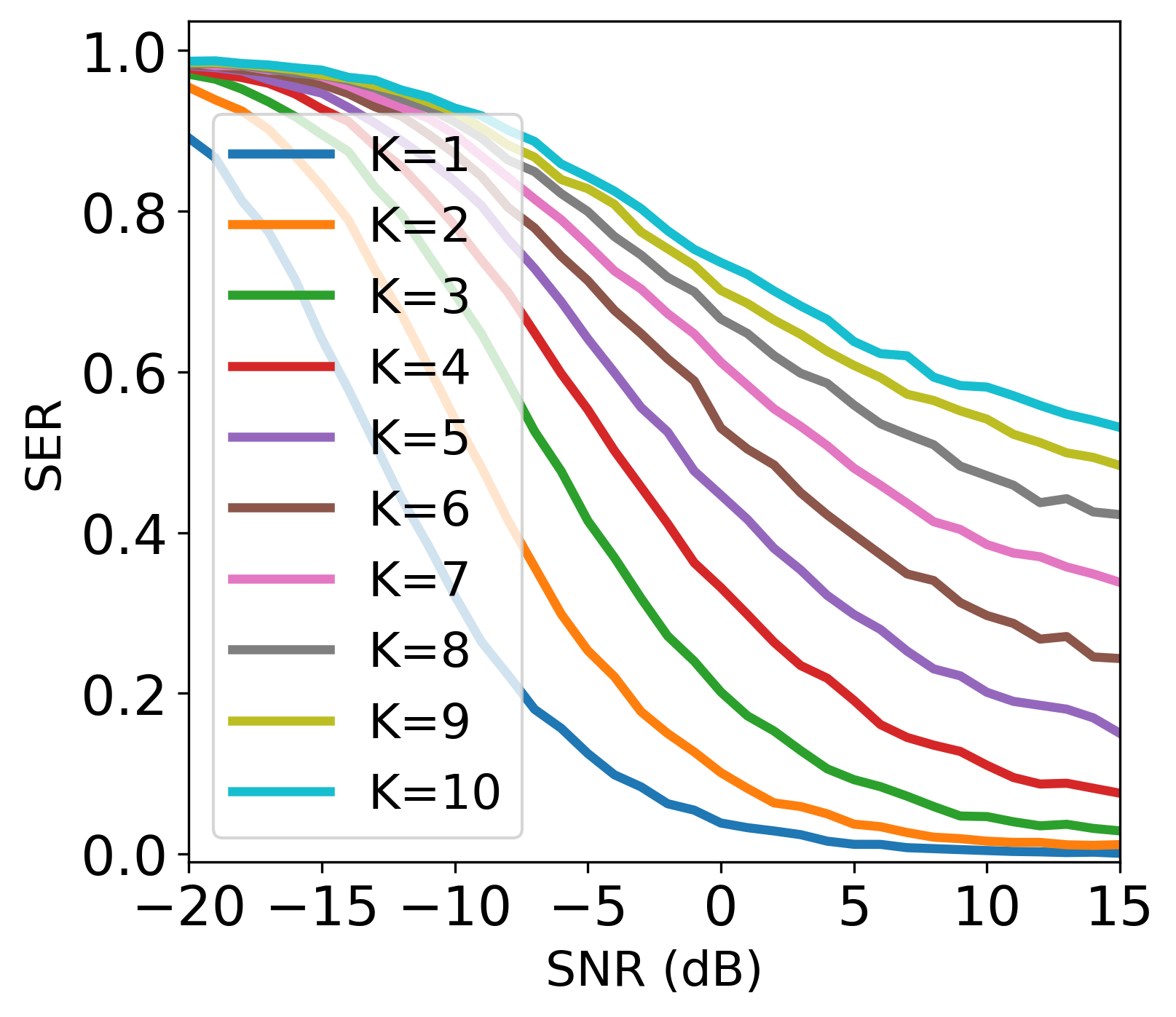}\label{fig:SER_VS_SNR_SF7}}
	\subfigure[]
	{\includegraphics[width=0.49\linewidth]{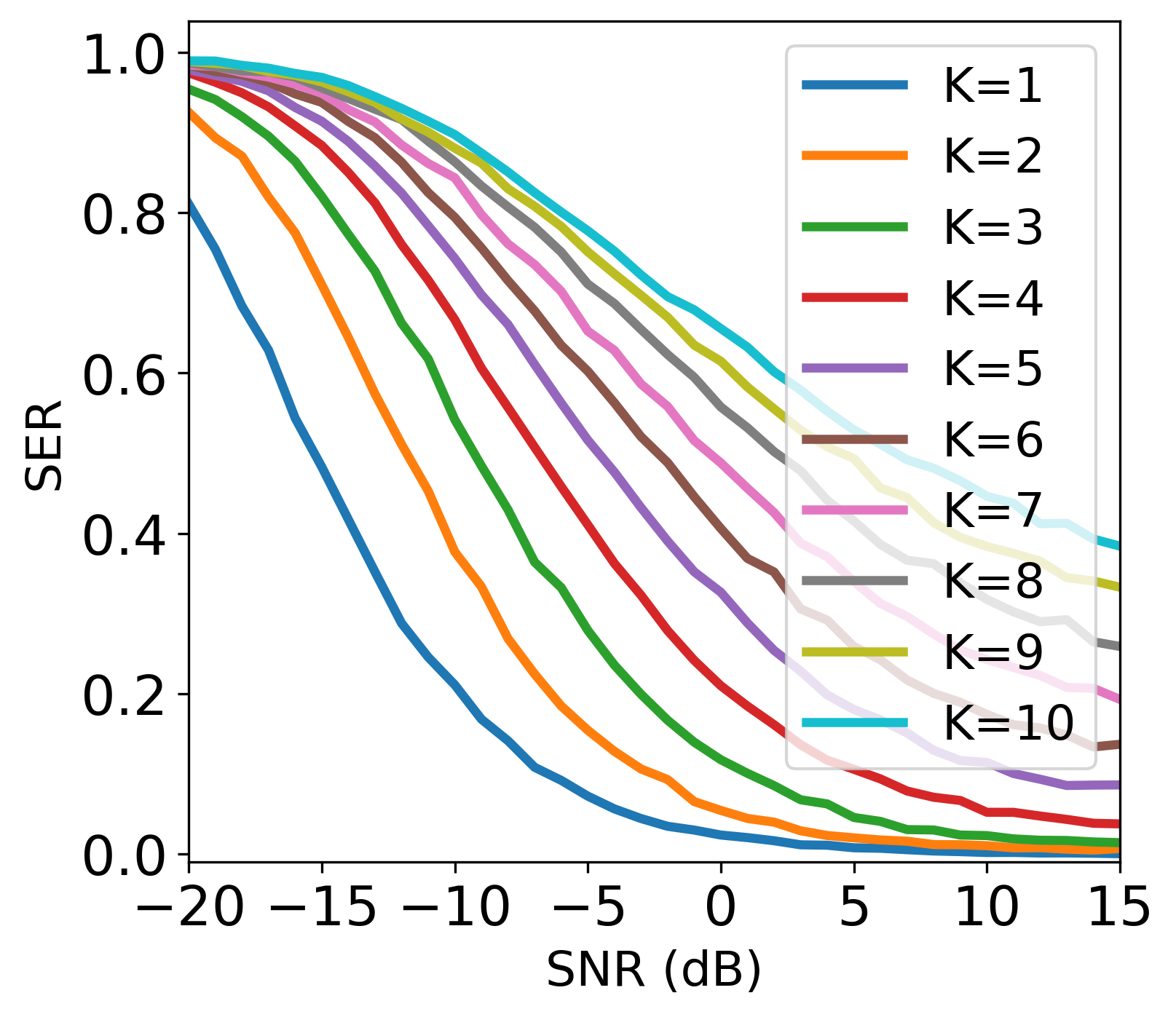}\label{fig:SER_VS_SNR_SF8}}
	\subfigure[]
	{\includegraphics[width=0.49\linewidth]{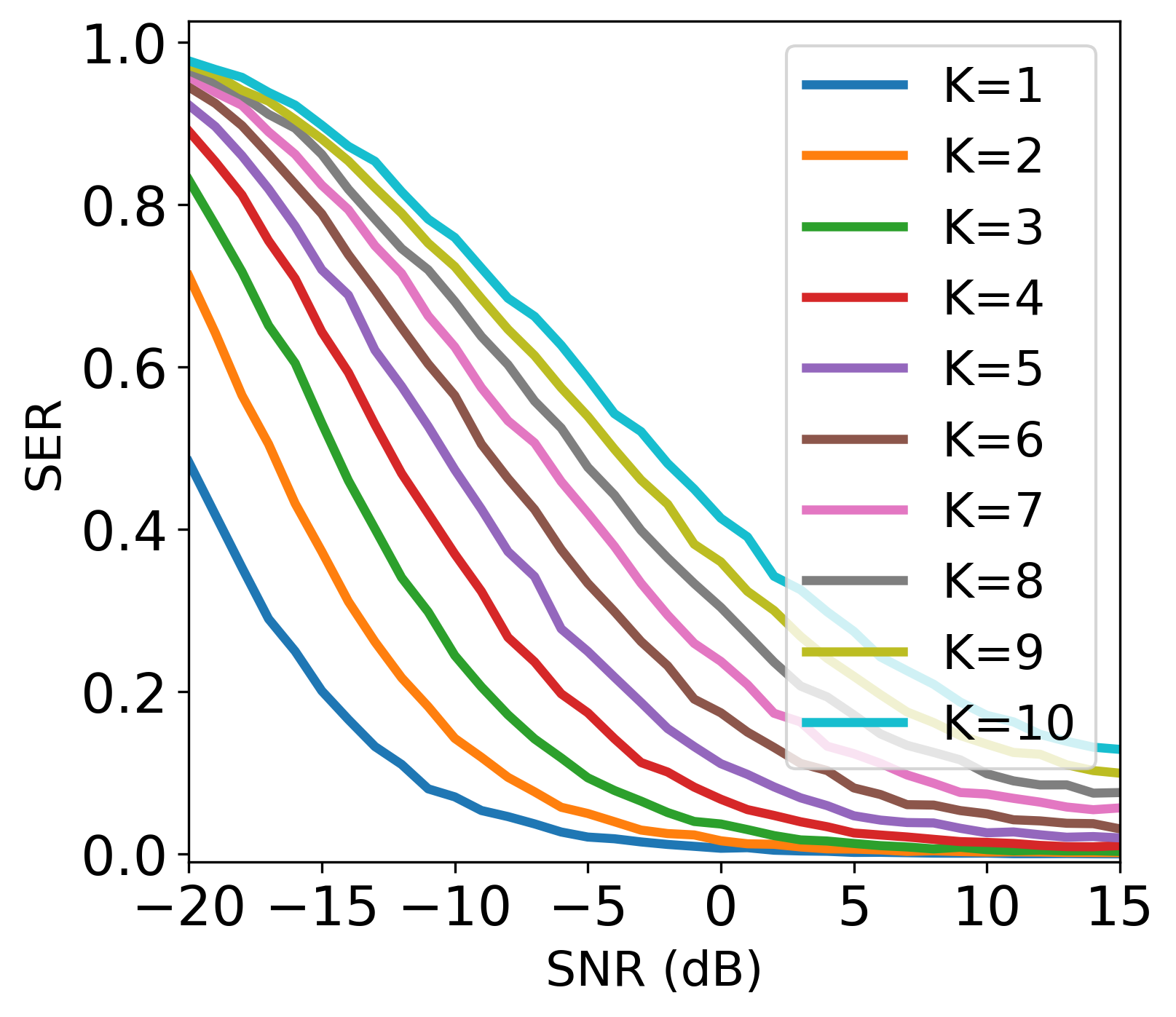}\label{fig:SER_VS_SNR_SF10}}
	\subfigure[]
	{\includegraphics[width=0.49\linewidth]{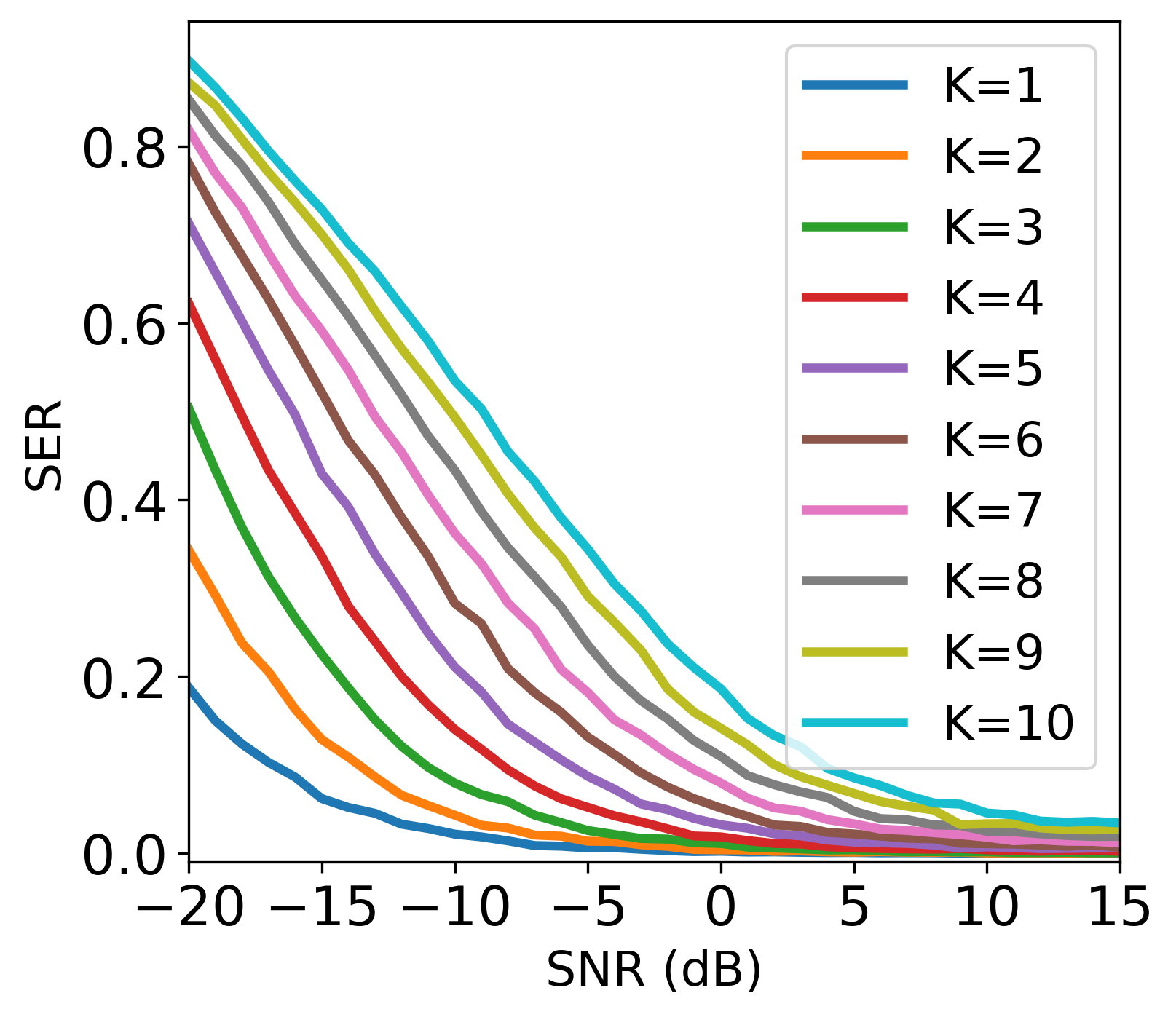}\label{fig:SER_VS_SNR_SF12}}	
	\caption{SER for\textit{ Super-LoRa} over \textcolor{black}{Rayleigh fading channel} with various Superposisiton orders and (a) $SF = 7$, (b) $SF = 8$, (c) $SF = 10$, and (d) $SF = 12$.}
	\label{fig:SER_VS_SNR_SF}
\end{figure}

We start by examining the SER as a function of SNR for various superposition orders $K$. The results are presented in Fig. \ref{fig:SER_VS_SNR_SF}.  The results reveal several insights, which we detail in the following. For SF7 (Fig. \ref{fig:SER_VS_SNR_SF7}), the framework achieves reliable performance ($SER \leq 10^{-1}$) for $K=2$ while maintaining an SNR around $0$ dB. 
This proves that we could expect the bitrate to be doubled even if the system is running at SNR lower than $0$ dB. For higher SNRs, the system can sustain higher superposition orders (up to $K = 4$) while maintaining $SER \leq 10^{-1}$. 
Further SNR would not improve the performance drastically as the signal would be impacted by the interfering symbols more than the noise.
This reveals that there is room to push the envelope of LoRa bitrate even at lower SNRs. Indeed, higher SF (similar to Figs. \ref{fig:SER_VS_SNR_SF8}, \ref{fig:SER_VS_SNR_SF10}, and \ref{fig:SER_VS_SNR_SF12}) can support higher superposition orders even at lower SNRs (lower than $0$ SNR for SF8, and lower than $-5$ SNR for SF10 and SF12). 
This stems from the inherent resilience of higher SF against noise and interference, hence, it allows more symbols to be transmitted concurrently. For example, with SF10,
three payload symbols can be superimposed and transmitted concurrently at $SNR \leq-5$
dB while maintaining $SER \leq 10^{-1}$. 
\par
\begin{figure}[h!]
	\centering
	\subfigure[]
	{\includegraphics[width=0.49\linewidth]{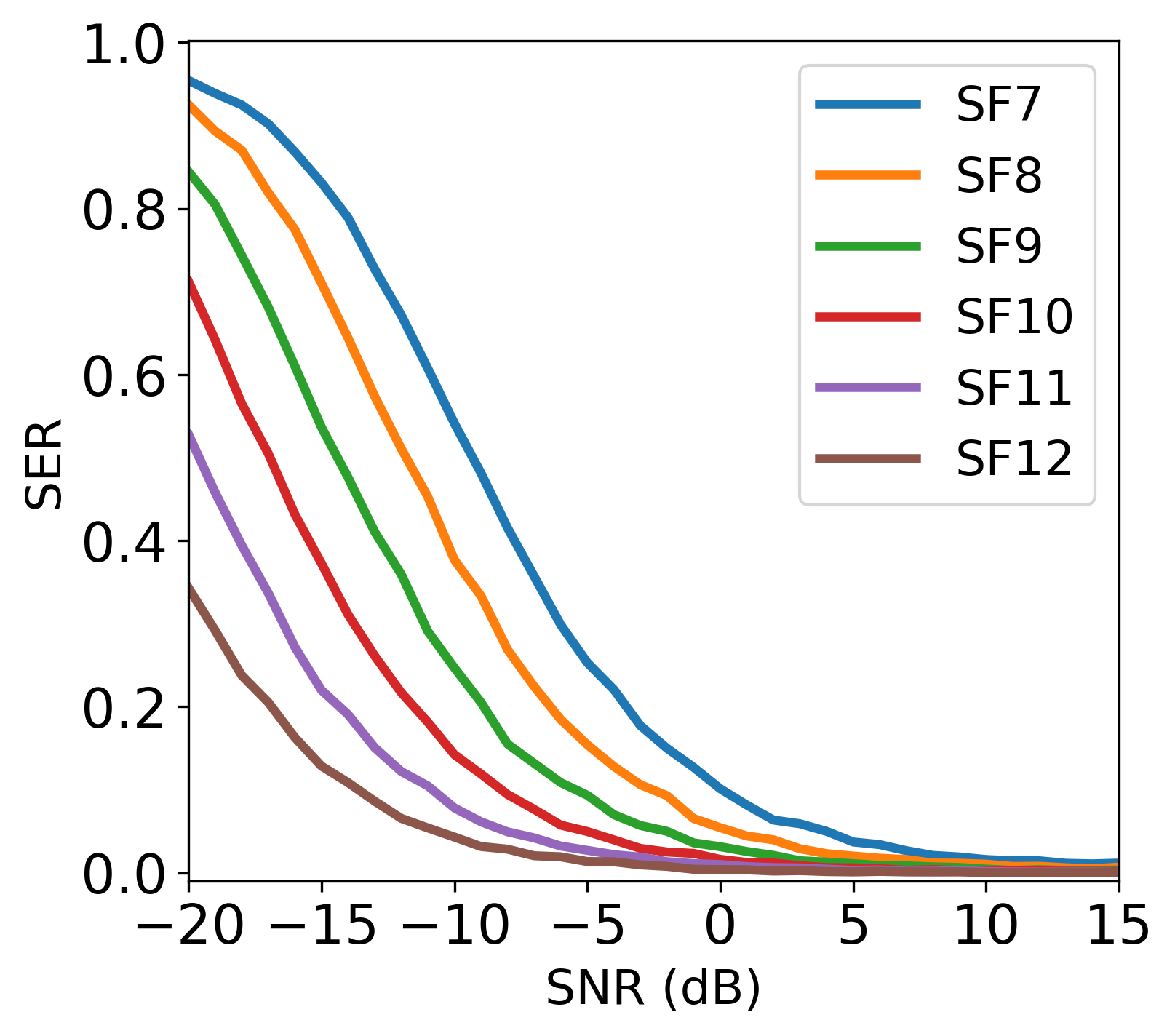}\label{fig:SER_VS_SNR_K2}}
	\subfigure[]
	{\includegraphics[width=0.49\linewidth]{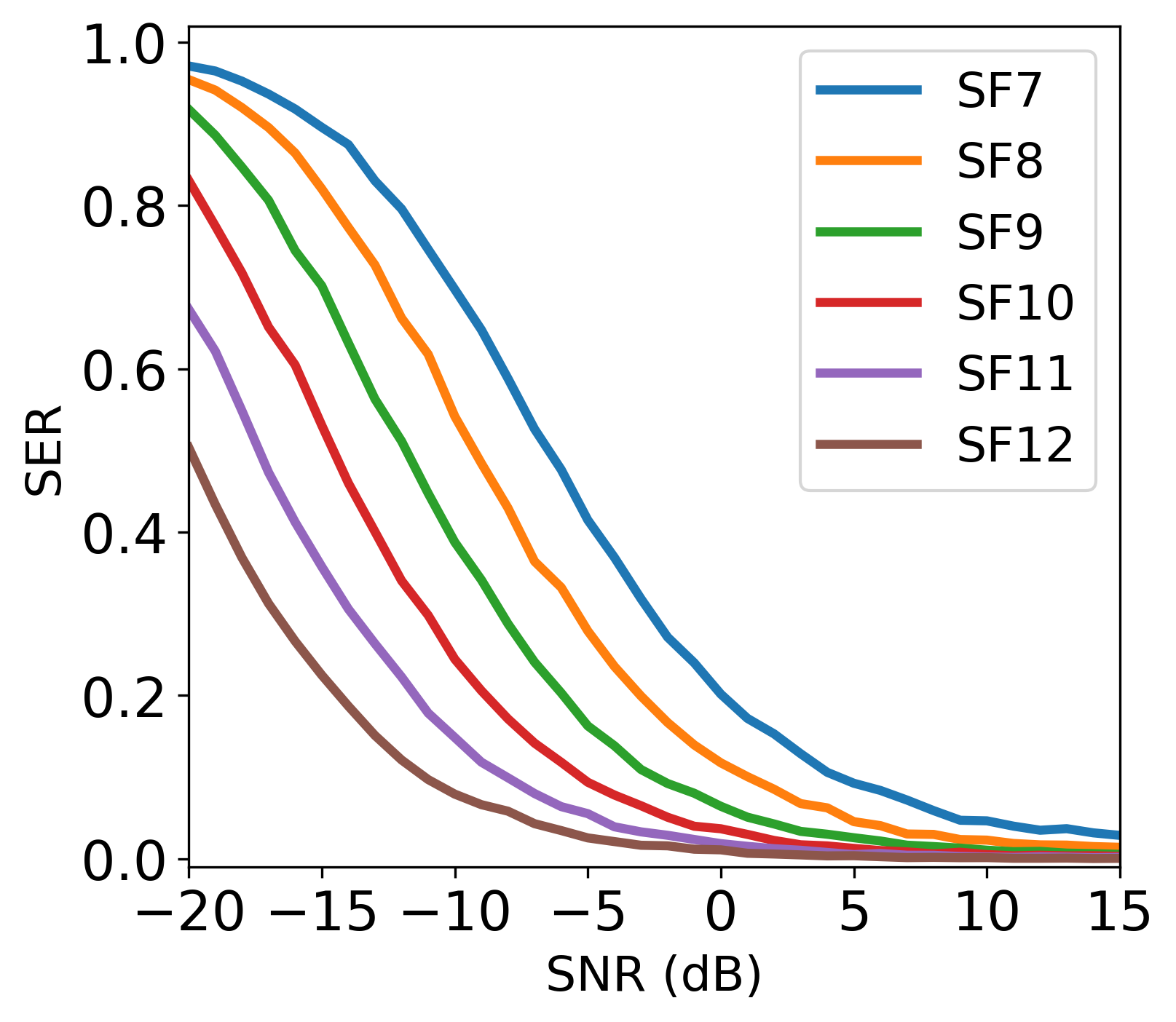}\label{fig:SER_VS_SNR_K3}}
	\subfigure[]
	{\includegraphics[width=0.49\linewidth]{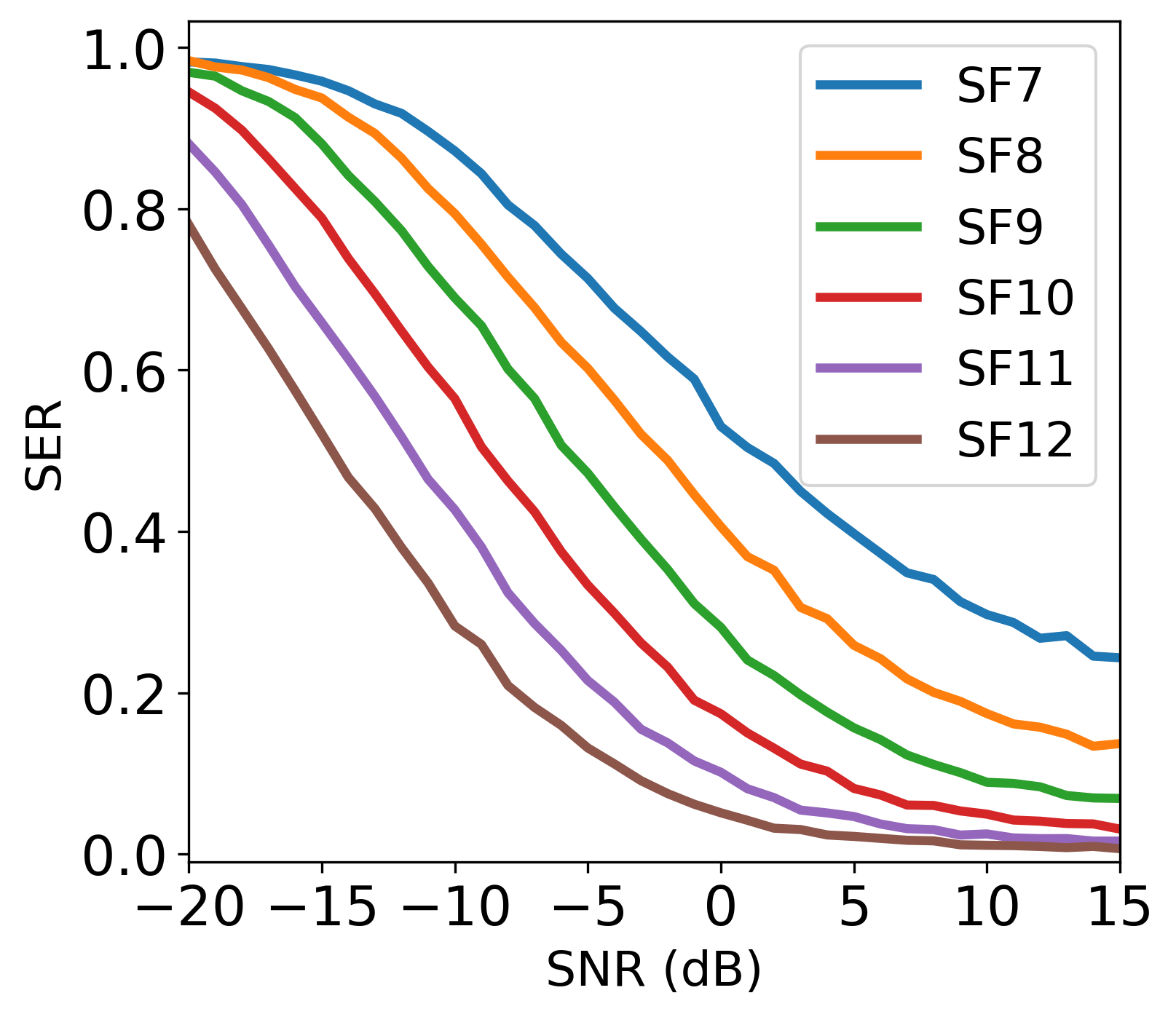}\label{fig:SER_VS_SNR_K6}}
	\subfigure[]
	{\includegraphics[width=0.49\linewidth]{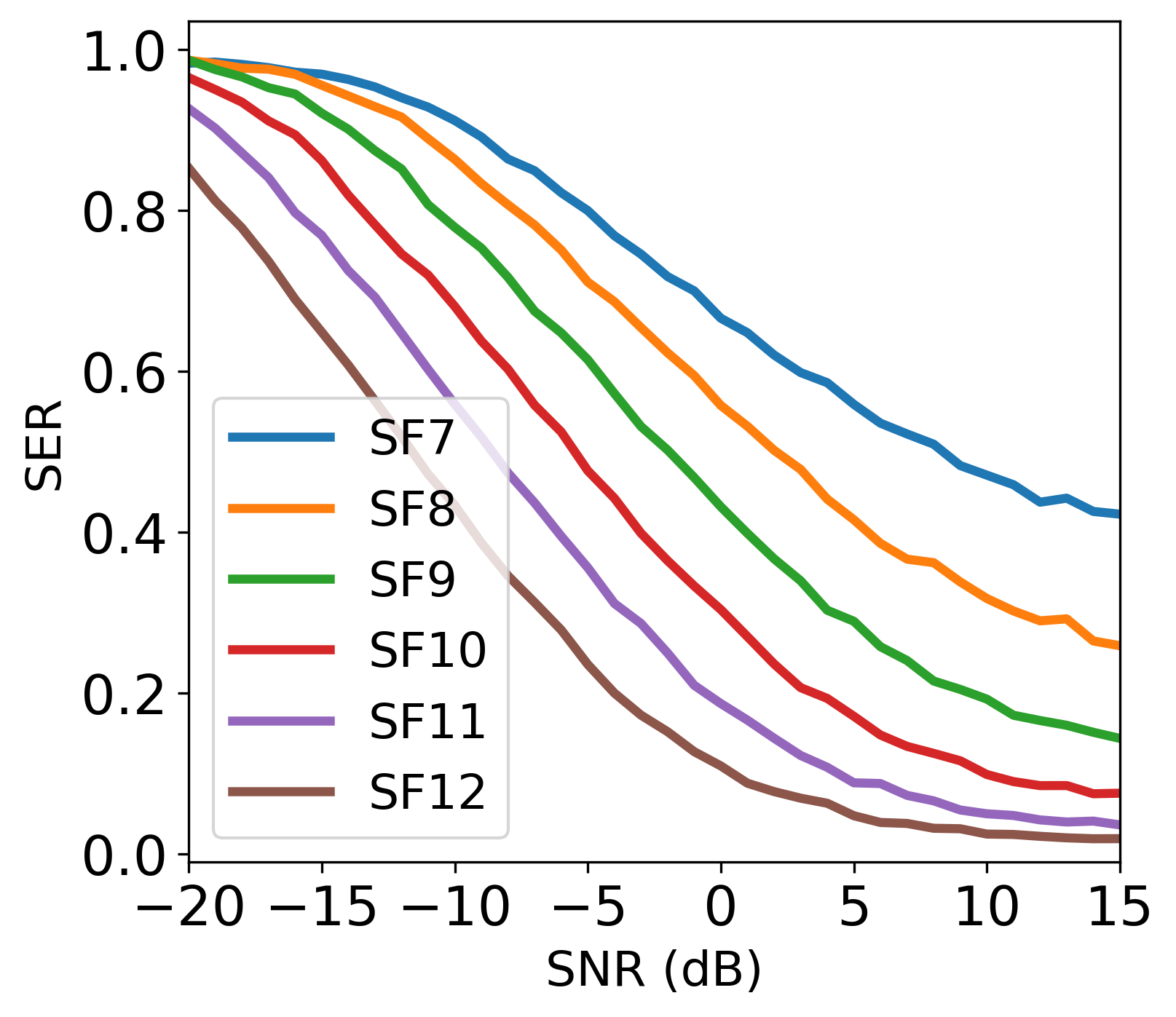}\label{fig:SER_VS_SNR_K8}}	
	\caption{SER for \textit{Super-LoRa} over \textcolor{black}{Rayleigh fading channel} with various SF and (a) $K = 2$, (b) $K = 3$, (c) $K = 6$, and (d) $K = 8$.}
	\label{fig:SER_VS_SNR_K}
\end{figure}

Similar observations can be deduced from Fig. \ref{fig:SER_VS_SNR_K}, where we plot SER vs SNR for various SFs. In Fig. \ref{fig:SER_VS_SNR_K2}, we can see that for superposition order $K =2$, we could reach a reliable communication ($SER \leq 10^{-1}$) at SNR around $0$ dB for all SF configurations. Similarly for $K =3$ shown in Fig. \ref{fig:SER_VS_SNR_K3}, $SER \leq 10^{-1}$ is reachable for SF10, SF11, and SF12 at $SNR \leq -5$ while it is reachable for SF8 at SNR around $1$ dB. A similar pattern can be observed for higher superposition orders, $K =6$ and $K =8$ shown in Fig. \ref{fig:SER_VS_SNR_K6} and Fig. \ref{fig:SER_VS_SNR_K8}, respectively, where higher superposition orders can be adopted at lower SNR ($SNR \leq -5$ dB) while still maintaining a reliable communication ($SER \leq 10^{-1}$).
\par
\begin{figure}[h!]
	\centering
	\subfigure[]
	{\includegraphics[width=0.49\linewidth]{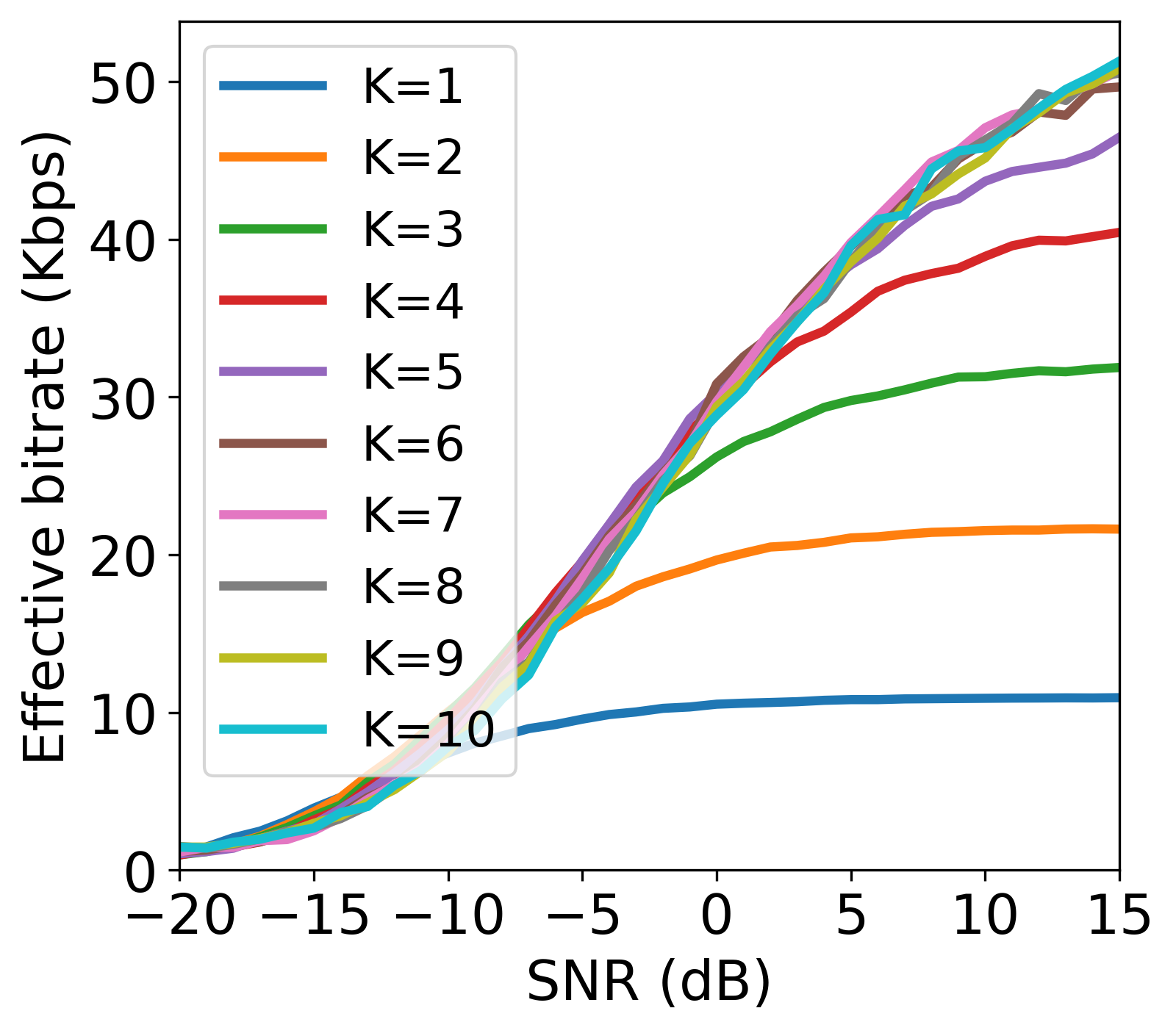}\label{fig:Rate_VS_SNR_BW250_SF7}}
	\subfigure[]
	{\includegraphics[width=0.49\linewidth]{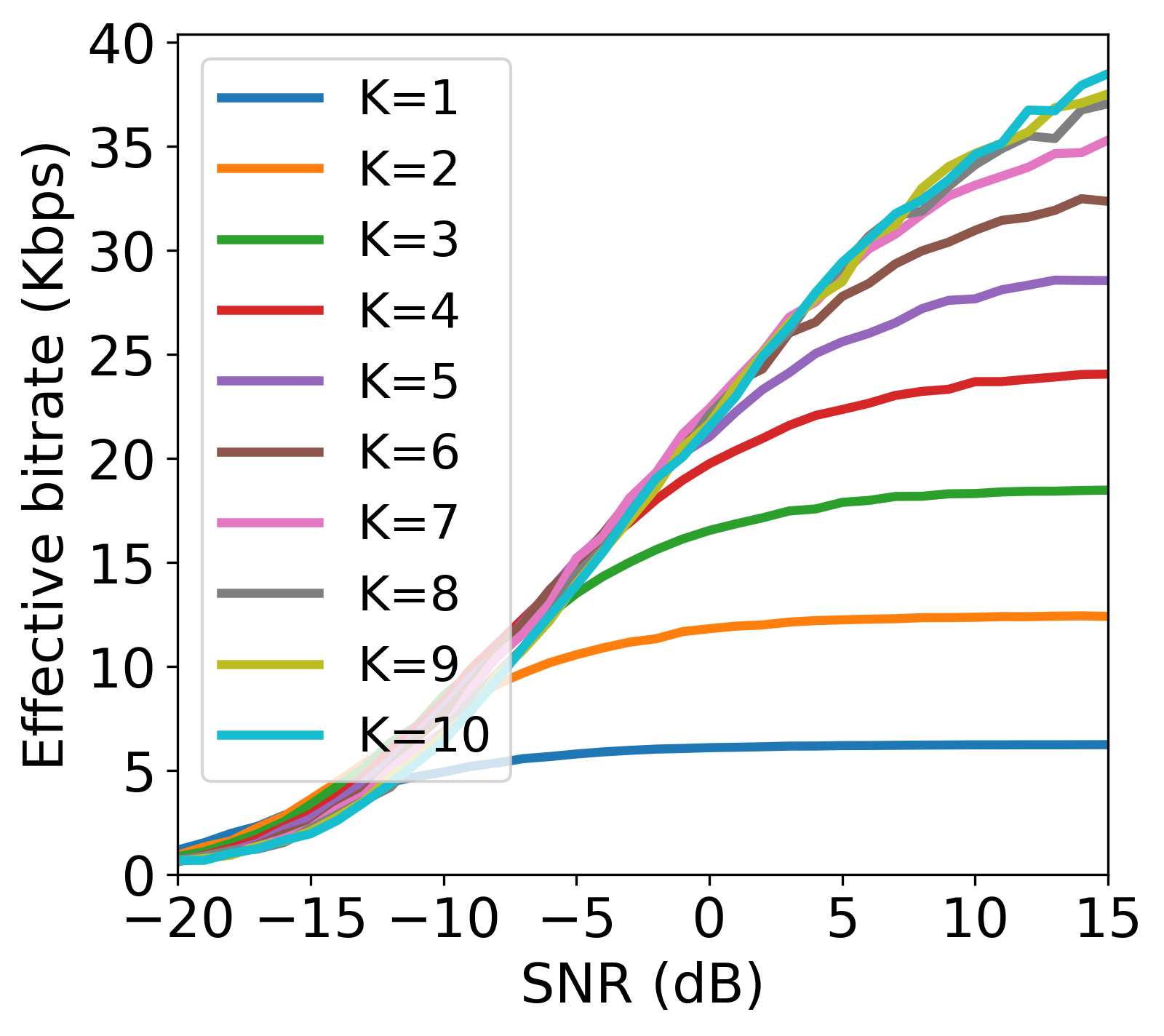}\label{fig:Rate_VS_SNR_BW250_SF8}}
	\subfigure[]
	{\includegraphics[width=0.49\linewidth]{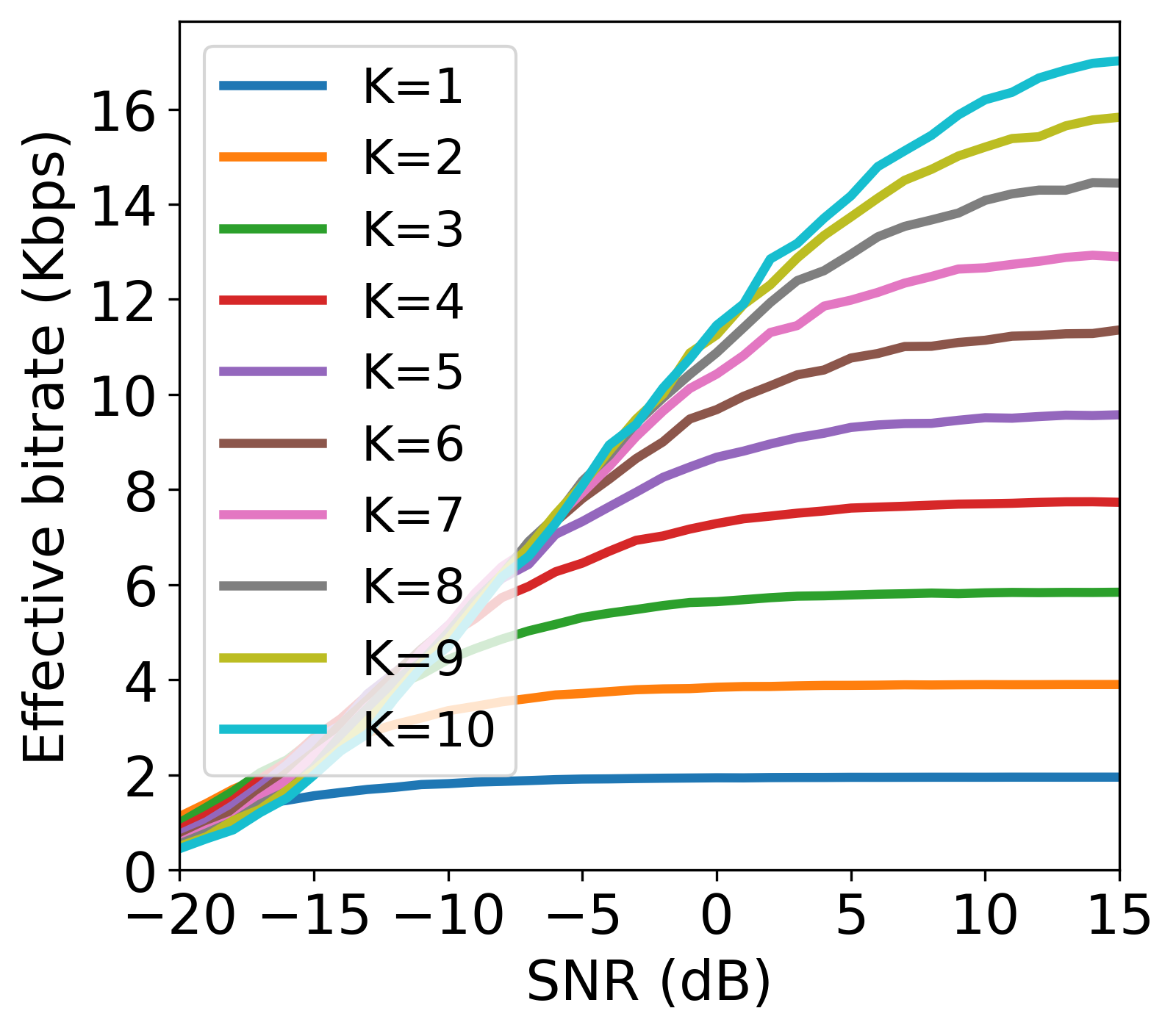}\label{fig:Rate_VS_SNR_BW250_SF10}}
	\subfigure[]
	{\includegraphics[width=0.49\linewidth]{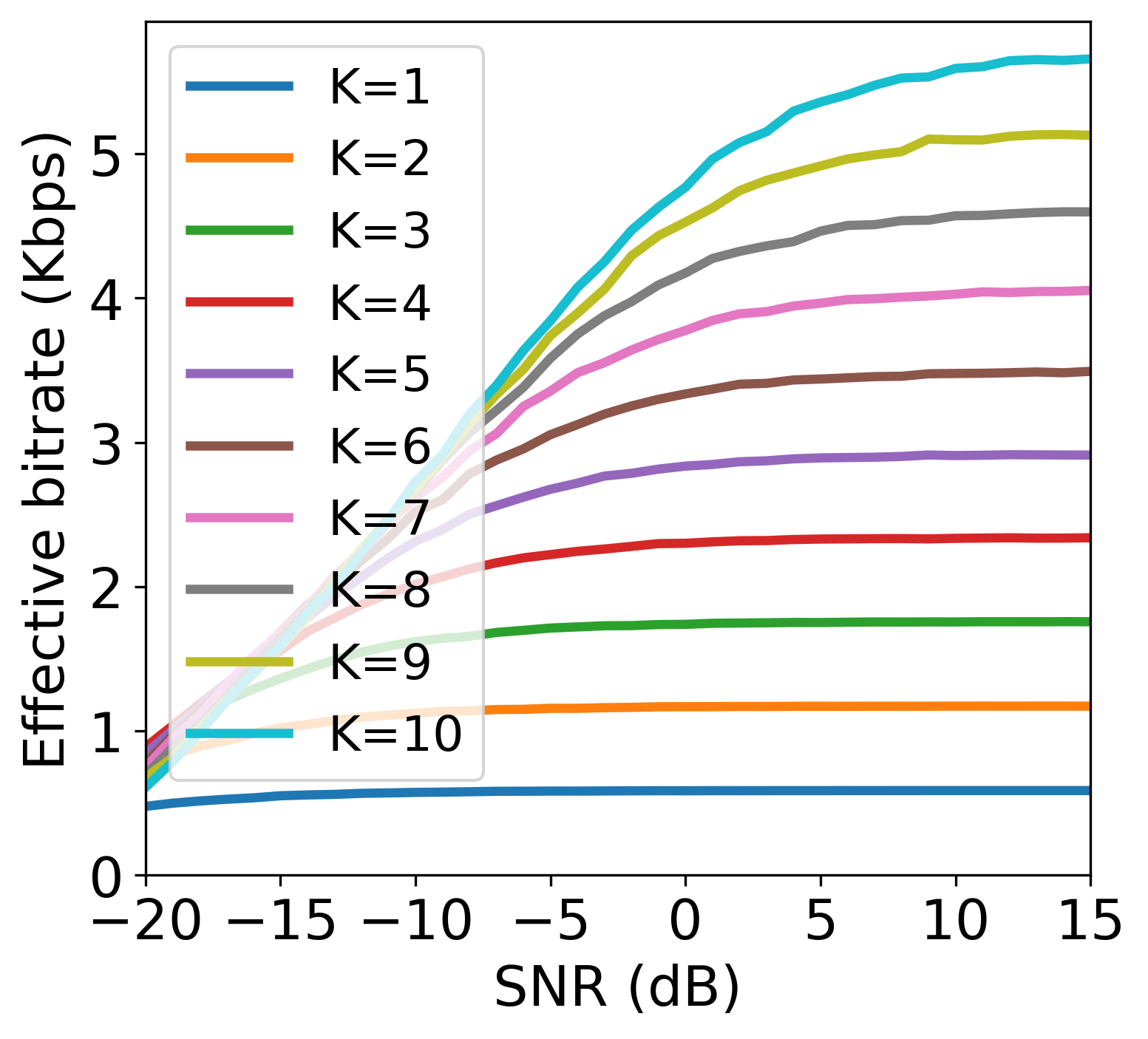}\label{fig:Rate_VS_SNR_BW250_SF12}}	
	\caption{Effective bitrate calculated based on Equ. \eqref{equ:effective_bitrate} for Super-LoRa for various superposition orders and (a) $SF = 7$, (b) $SF = 8$, (c) $SF = 10$, and (d) $SF = 12$.}
	\label{fig:Rate_VS_SNR_BW250_SF}
\end{figure}

Finally, we present in Fig. \ref{fig:Rate_VS_SNR_BW250_SF} the effective bitrate calculated with Equ. \eqref{equ:effective_bitrate} as a function of SNR for different SF configurations. Starting with SF7 {Fig. \ref{fig:Rate_VS_SNR_BW250_SF7}}, it is shown that with SNR around $-5$ dB, the system would have bitrate around $20$ Kbps with $K=2$ compared to $10$ Kbps for standard LoRa (with $K=1$). It clearly shows the superior improvement of bitrate at lower SNR. The bitrate can be pushed further at higher SNR (SNR $= 0$ dB) to reach around $30$ Kbps for $K=3$. This shows the great potential of \textit{Super-LoRa} to improve the system’s bitrate. Further bitrate improvement can be seen at higher SNR to reach around $40$ Kbps with $K = 4$ at SNR$\geq 5$ dB. Similar improvement can be seen with SF8, SF10, and SF12 in Figs. \ref{fig:Rate_VS_SNR_BW250_SF8}, \ref{fig:Rate_VS_SNR_BW250_SF10}, and \ref{fig:Rate_VS_SNR_BW250_SF12}, where the bitrate is improved even at lower SNRs (SNR $\leq -10$ dB).

\subsection{Experimental Setup}
\par
\textit{Super-LoRa} was implemented on Software Defined Radios (SDR) using the \textit{USRP N210} \cite{usrp_n210} and the low-cost \textit{ADALM-PLUTO} SDR \cite{adalm_pluto}. The modulator and demodulator were integrated as new blocks into the popular SDR software, \textit{GNU Radio} \cite{gnu_radio}. Since the transmitter and receiver need to remain LoRa-compliant, we used \textit{\textbf{gr-lora\_sdr}} \cite{tapparel2020open}, an open-source GNU Radio LoRa implementation. We modified the \textit{\textbf{modulator}} block to accept a new parameter, superposition order $K$, and perform the superposition as described in Equ. \eqref{equ:SuperPosition_Modulation}. Additionally, we adjusted the \textit{\textbf{Frame sync}} block in the demodulator to accept the same superposition order $K$, allowing frame demodulation every $\tau$ seconds instead of $T_s$. Notably, the encoder and decoder components remained unchanged, requiring only additional buffering at both the modulator and demodulator, without extra processing overhead as described in sections \ref{Super-LoRa-Modulator} and \ref{Super-LoRa-Demodulator}.
\par
\begin{figure}[h!]
	\centering
	\subfigure[]
	{\includegraphics[width=0.63\linewidth]{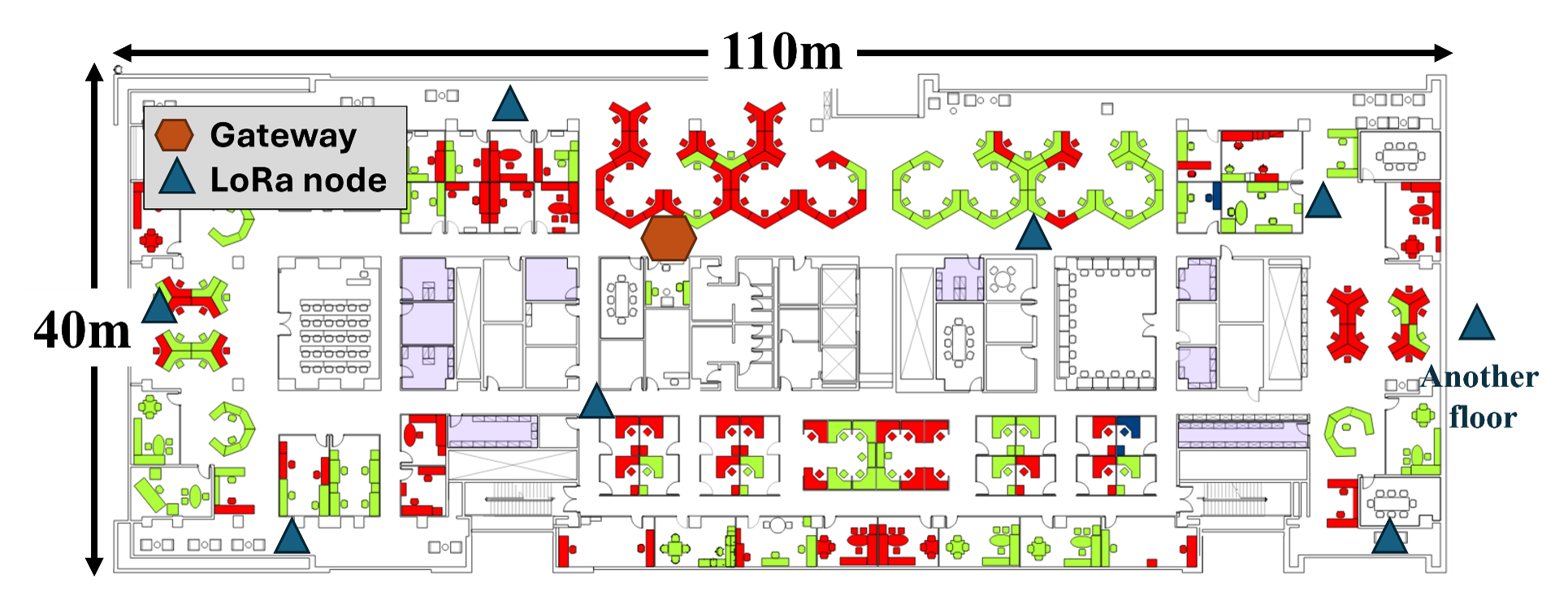}\label{fig:indoor}}
	\subfigure[]
	{\includegraphics[width=0.34\linewidth]{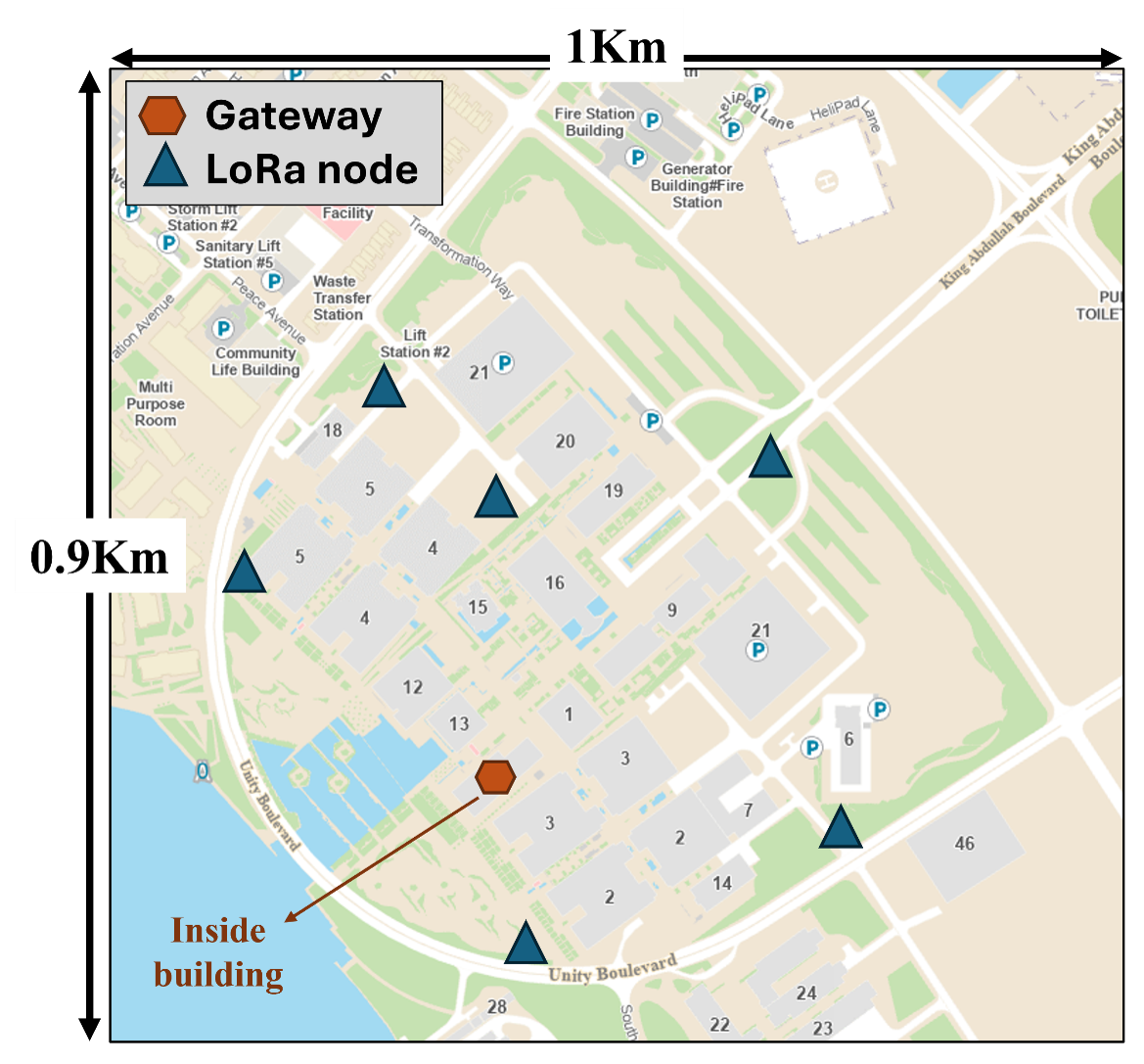}\label{fig:outdoor}}
 	\subfigure[]
	{\includegraphics[scale=0.2]{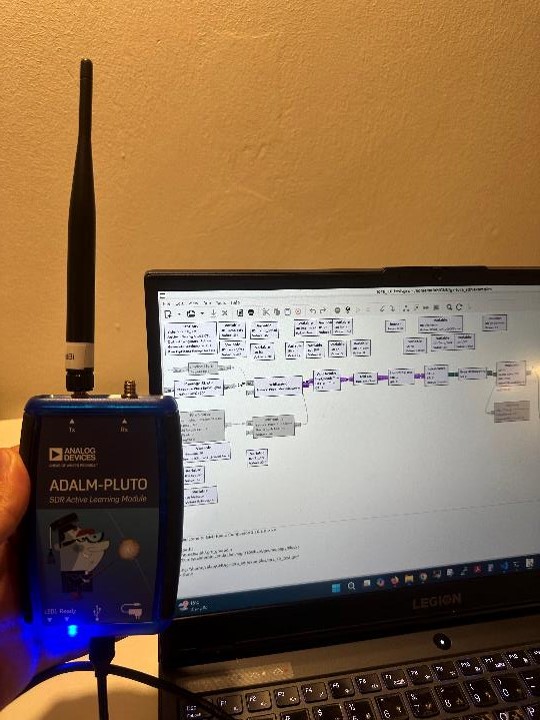}\label{fig:TX_ADALM}}
	\subfigure[]
	{\includegraphics[scale=0.2]{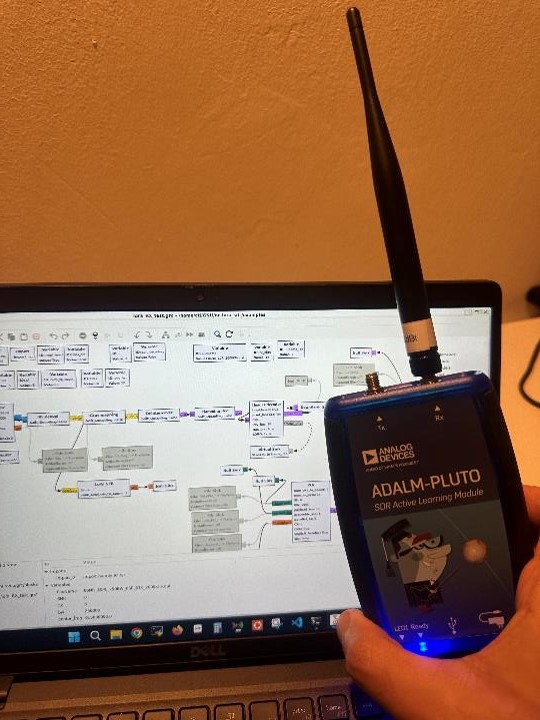}\label{fig:RX_ADALM}}
 \caption{Deployment testbeds of \textit{Super-LoRa} on (a) indoor, and (b) outdoor environments. \textit{Super-LoRa} implementation on \textit{GNU Radio} for the (c) transmitter, and (d) receiver.}\label{fig:Deplyment_Testbed}
\end{figure}

Over the course of a month, we collected transmission traces of \textit{Super-LoRa} across our university campus using various transmission parameters and link conditions. The collected traces and node locations are depicted in Fig. \ref{fig:Deplyment_Testbed}. The setup spans a $1 \, \text{Km} \times 0.9 \, \text{Km}$ area, covering both indoor and outdoor environments with Line-of-Sight (LoS) and non-LoS transmissions. \textit{Super-LoRa} operated on the 915 MHz ISM band with a bandwidth of 250 KHz. The links encountered various SNR conditions, ranging from $< -5 \, \text{dB}$ to $> 5 \, \text{dB}$.
\par
It should be noted that \textit{Super-LoRa} cannot be directly integrated with Commercial-Off-the-Shelf (COTS) LoRa nodes, as access to the I-Q samples of the modulator/demodulator is required. However, as discussed in Section \ref{Design_Consideration}, we reused all components of the encoder and decoder, only modifying the modulator and demodulator.

\subsubsection*{\textbf{SER Vs SNR}}
\par
\begin{figure}[h!]
	\centering
	\subfigure[]
	{\includegraphics[width=0.49\linewidth]{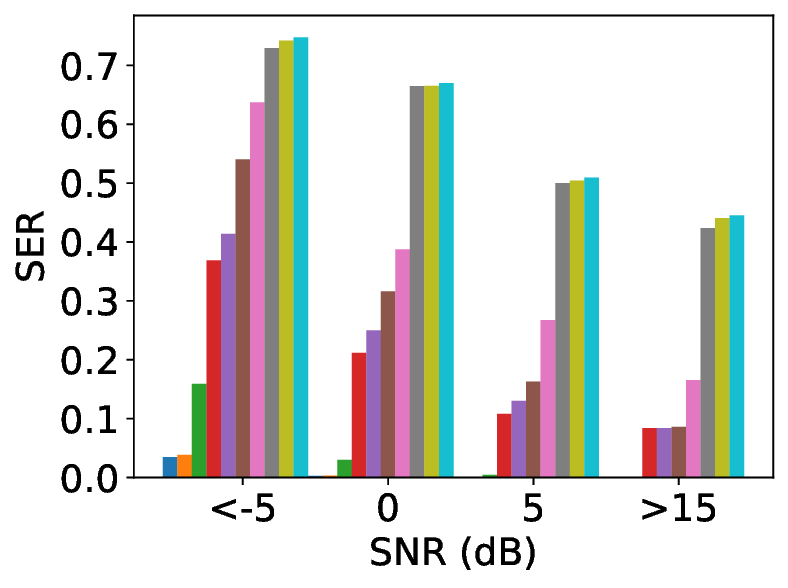}\label{fig:EXP_SER_VS_SNR_SF7}}
	\subfigure[]
	{\includegraphics[width=0.49\linewidth]{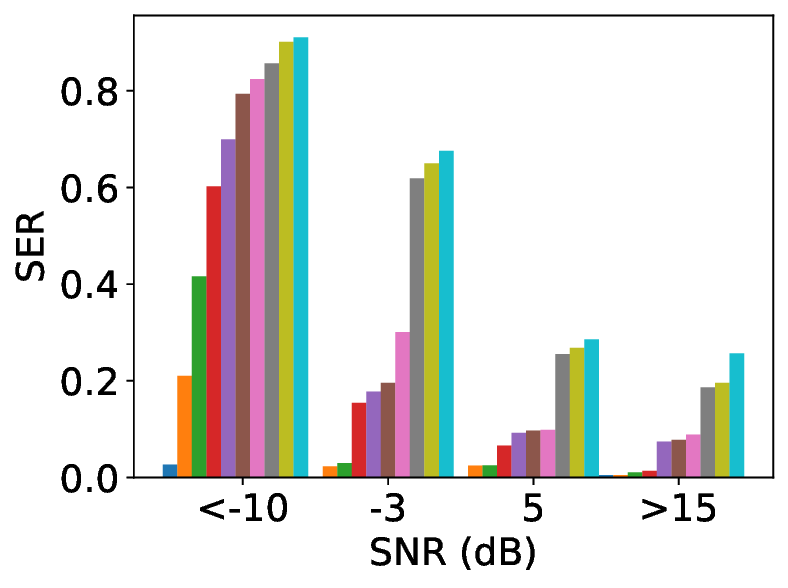}\label{fig:EXP_SER_VS_SNR_SF8}}
	\subfigure[]
	{\includegraphics[width=0.8\linewidth]{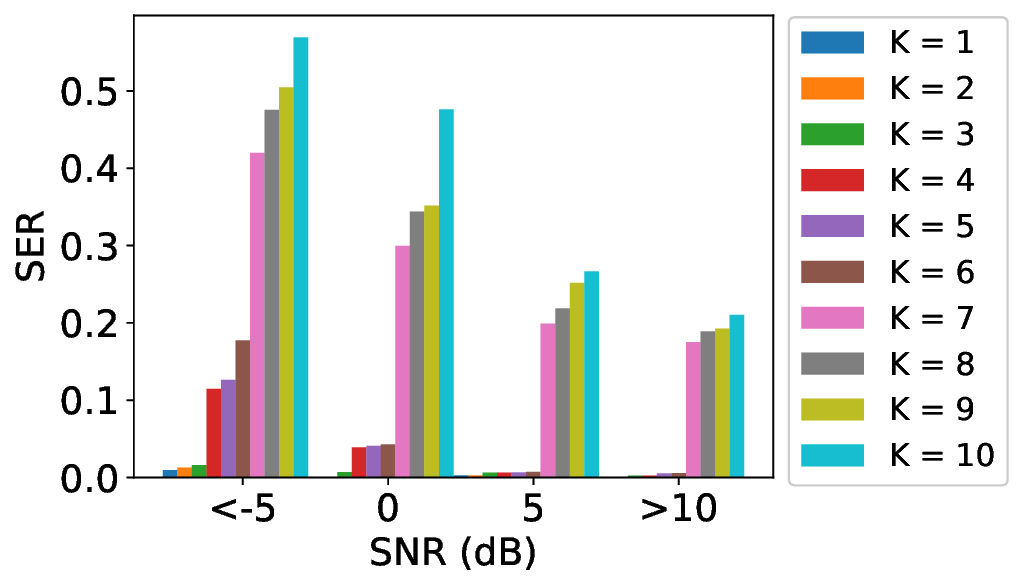}\label{fig:EXP_SER_VS_SNR_SF9}}
	\caption{SER for \textit{Super-LoRa} for various superposition orders and (a) $SF = 7$, (b) $SF = 8$, and (c) $SF = 9$.}
	\label{fig:EXP_SER_VS_SNR_SF}
\end{figure}

We evaluate the SER of \textit{Super-LoRa} through experiments for SF7, SF8, and SF9, plotting SER as a function of SNR for different superposition orders $K$ (Fig. \ref{fig:EXP_SER_VS_SNR_SF}). For SF7 (Fig. \ref{fig:EXP_SER_VS_SNR_SF7}), reliable performance (SER $\leq 10^{-1}$) is achieved at $K=2$ and SNR $\leq -5$ dB, demonstrating that the bitrate can be improved even at low SNRs. For higher SNRs, $K = 5$ can be supported while maintaining SER $\leq 10^{-1}$. Higher SF values (Figs. \ref{fig:EXP_SER_VS_SNR_SF8}, \ref{fig:EXP_SER_VS_SNR_SF9}) allow higher superposition orders at similarly low SNRs, such as $K=3$ at SNR $\leq -5$ dB with SF9.

\subsubsection*{\textbf{Link Throughput}}
\par
\begin{figure}[h!]
	\centering
	\subfigure[]
	{\includegraphics[width=0.49\linewidth]{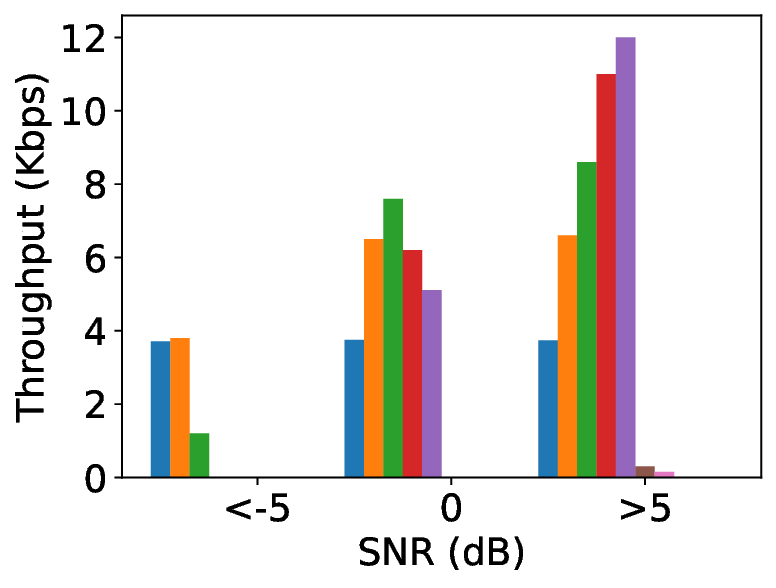}\label{fig:EXP_Rate_VS_SNR_SF7}}
	\subfigure[]
	{\includegraphics[width=0.49\linewidth]{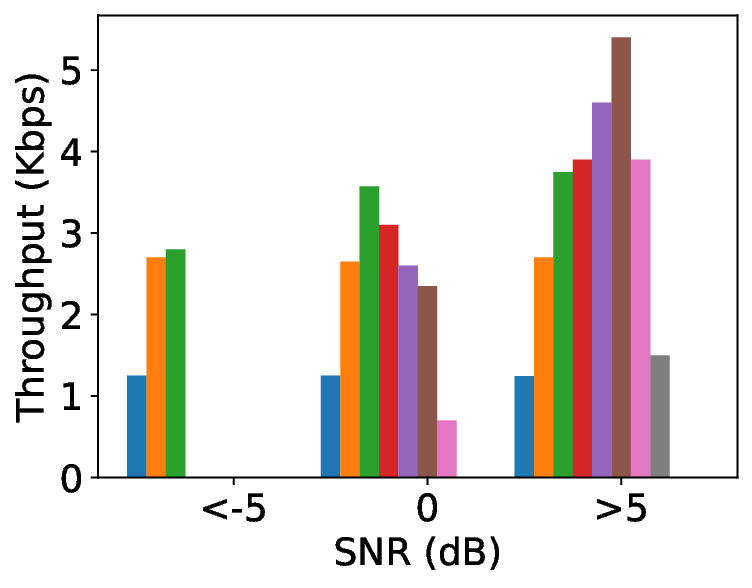}\label{fig:EXP_Rate_VS_SNR_SF8}}
	\subfigure[]
	{\includegraphics[width=0.8\linewidth]{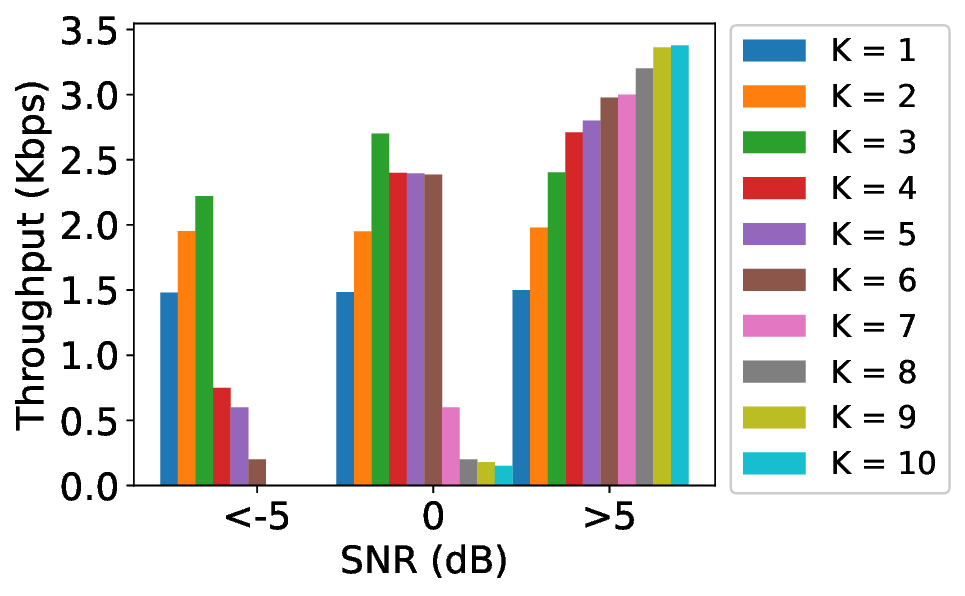}\label{fig:EXP_Rate_VS_SNR_SF9}}
	\caption{Throughput for Super-LoRa for various Superposition orders $K$ and (a) $SF = 7$, (b) $SF = 8$, and (c) $SF = 9$.}
	\label{fig:EXP_Rate_VS_SNR_SF}
\end{figure}

We evaluate the throughput improvement of \textit{Super-LoRa} under various $SNR$ conditions by transmitting a 10-byte payload with different SF, superposition order $K$, and channel conditions (i.e., SNR). Note that the packet overhead (header, CRC, etc.) is the same for all superposition orders $K$ as \textit{Super-LoRa} only superimpose the payload symbols. The results are depicted in Fig. \ref{fig:EXP_Rate_VS_SNR_SF}. For SF7 (Fig. \ref{fig:EXP_Rate_VS_SNR_SF7}), a throughput of $4$ Kbps is observed for $K = 1$ at SNR $\leq -5$. However, at SNR around $0$ dB, the throughput increases to $6$ Kbps and $8$ Kbps for $K=2$ and $K=3$, respectively, showing a $1.5\times$ to $2\times$ improvement over standard LoRa. The improvement extends up to $3\times$ at higher SNR (SNR $\geq 5$), reaching $12$ Kbps. This shows the great potential of \textit{Super-LoRa} to improve the communication throughput. Similar observations are seen for higher SF, as depicted in Fig. \ref{fig:EXP_Rate_VS_SNR_SF8} and Fig. \ref{fig:EXP_Rate_VS_SNR_SF9}, where the improvement extends up to $5\times$ compared to standard LoRa at higher SNR regime, highlighting the potential of \textit{Super-LoRa} in data-intensive IoT applications.

\subsection{Discussion}
\par
While the previous sections highlighted the improvements offered by \textit{Super-LoRa} across various SF configurations, the most significant gains are observed at the lowest SF (\(SF = 7\)). This is particularly relevant since LoRa devices with higher SNRs are better suited to operate at lower SFs. Nonetheless, \textit{Super-LoRa} can also effectively exploit SNRs near the transition thresholds between different SFs, such as those used in ADR in LoRaWAN networks.
\par
\textcolor{black}{In addition, it is worth mentioning that  
the most relevant comparison to our work is HyLink \cite{xia2022hylink}, which utilizes amplitude and phase signatures of LoRa symbols to transmit them concurrently. While their approach achieves notable throughput improvements, their implementation involves higher computational complexity to retrieve the amplitude and phase signatures. Further, since their implementation is not publicly available, making a direct comparison with our framework is challenging. However, based on reported SER and throughput metrics in \cite{xia2022hylink}, our framework achieves comparable performance under similar conditions, with the added benefit for our framework of requiring minimal modifications to the standard LoRa architecture.}

\section{related Works}
\par
Many previous works have focused on concurrent decoding of collided LoRa packets at LoRa gateway to improve the overall network capacity, including \textit{Choir} \cite{eletreby2017empowering}, \textit{mLoRa} \cite{wang2019mlora}, \textit{CIC} \cite{shahid2021concurrent}, \textit{CoLoRa} \cite{tong2020colora}, \textit{NScale} \cite{tong2020combating}, and \textit{FTrack} \cite{xia2019ftrack}. \textit{Choir} \cite{eletreby2017empowering} leverages the hardware imperfections that produce unique CFO signatures for different LoRa transmitters, allowing the classification of collided packets based on frequency offsets. \textit{FTrack} \cite{xia2019ftrack} applies Short Time Fourier Transform (STFT) on dechirped LoRa symbols, utilizing time and frequency features to enhance collision resolution by leveraging spread spectrum gain and removing linear frequency variations. \textit{CIC} \cite{shahid2021concurrent} accumulates windows of varying lengths in time and frequency domains, balancing resolution to decode collided packets. \textit{NScale} \cite{tong2020combating} decodes LoRa collisions at low SNR by converting timing offsets into frequency features and amplifying time offsets through non-stationary scaling. \textit{mLoRa} \cite{wang2019mlora} and \textit{CoLoRa} \cite{tong2020colora} resolve packet collision by classifying symbols based on the received power levels of each packet. 
Unlike these works, \textit{Super-LoRa} focuses on enhancing the communication throughput of individual LoRa links, supporting data-intensive IoT applications per each link.

The work most closely related to ours is \textit{HyLink} \cite{xia2022hylink}, which proposes an enhancement to LoRa networks called parallel Chirp Spread Spectrum (p-CSS) to increase data rates. HyLink modulates multiple symbols within a single chirp duration, identifying each symbol by encoding unique \textit{phase} and \textit{amplitude} signatures, enabling concurrent transmission. This approach relies on extracting these signatures at the receiver to demodulate parallel transmissions, especially in high-SNR environments. In contrast, our proposed \textit{Super-LoRa} framework employs pre-agreed \textit{time-delay} signatures between superimposed symbols, eliminating the need for signature extraction during demodulation. This design reduces receiver complexity while maintaining high throughput, setting our approach apart from HyLink’s reliance on phase and amplitude extraction.

\section{Conclusion}
\par
In this work, we introduced \textit{Super-LoRa}, an innovative transmission technique that pushes the boundaries of LoRa’s throughput while maintaining lower transmitter and receiver complexity. Through simulations and real-world evaluations, we demonstrated that \textit{Super-LoRa} can achieve up to a $5\times$ increase in throughput with minimal additional complexity at the receiver. Our results suggest that \textit{Super-LoRa} is a promising candidate for IoT applications requiring higher data rates, such as real-time environmental monitoring and smart city infrastructure. 
Future work will focus on optimizing the framework parameters, expanding compatibility with commercial LoRa devices, and \textcolor{black}{incorporating interference cancellation techniques to further improve the robustness of \textit{Super-LoRa.}}

\textcolor{black}{
\appendices
\section{\textbf{Optimal Power allocation proof}} \label{sec:Optimal_Power_Allocation_Proof}
We show in this section that uniform power allocation in \textit{Super-LoRa}, i,e, $P_1=P_2=\dots=P_K=\frac{P_T}{K}$, ensures optimum power allocation to maximize the minimum SIR across all superimposed payload symbols. Defining the power allocation vector as \( \boldsymbol{P} = \left[ P_1, P_2, \dots, P_K \right]^T \), an optimization problem can be formulated as:
\begin{align}
    \max_{\boldsymbol{P}} \quad & \min_{j} \left\{ \frac{P_j}{\sum_{\substack{i=1 \\ i \neq j}}^{K} P_i} \;\middle|\; j \in \{1, 2, \dots, K\} \right\}  \\
    \text{s.t.} \quad & \sum_{i=1}^{K} P_i = P_T.
\end{align}
To facilitate solving this problem, we introduce an auxiliary variable \( t \), replacing the minimum function. The problem then reformulates as:
\begin{align}
    \max_{t\boldsymbol{, P}} \quad & t  \label{Equ:objective_function_modified} \\
    \text{s.t.} \quad & \frac{P_j}{\sum_{\substack{i=1 \\ i \neq j}}^{K} P_i} \geq t, \quad \forall j \in \{1, 2, \dots, K\} \label{Equ:min_SIR_constraint} \\
    & \sum_{i=1}^{K} P_i = P_T. \label{Equ:sum_power_constraint}
\end{align}
To determine a feasible upper bound of the objective function, we express \( P_j \) from \eqref{Equ:sum_power_constraint} as:
\begin{equation}
    P_j = P_T - \sum_{\substack{i=1 \\ i\neq j}}^{K} P_i. \label{Equ:min_SIR_constraint_modified}
\end{equation}
Additionally, the constraint in \eqref{Equ:min_SIR_constraint} can be rewritten as:
\begin{equation}
    P_j \geq t \left( \sum_{\substack{i=1 \\ i\neq j}}^{K} P_i \right), \quad \forall j \in \{1, 2, \dots, K\}. \label{Equ:sum_power_constraint_modified}
\end{equation}
Since all \( P_j \) and \( t \) are positive, summing both sides of \eqref{Equ:sum_power_constraint_modified} over all \( j \) and combining it with \eqref{Equ:min_SIR_constraint_modified} yields:
\begin{equation}
    \sum_{i=1}^{K} P_i \geq t \left( (K - 1) \sum_{i=1}^{K} P_i \right) \Rightarrow t \leq \frac{1}{K-1}. \label{Equ:upper_bound}
\end{equation}
Equation \eqref{Equ:upper_bound} establishes an upper bound for the objective function in \eqref{Equ:objective_function_modified}. To verify achievability of the upper bound $t^* = \frac{1}{K-1}$, we consider the uniform power allocation:
\begin{equation}
    P_1 = P_2 = \dots = P_K = \frac{P_T}{K}.
\end{equation}
Substituting into the SIR expression:
\begin{equation}
    \frac{P_T/K}{(K-1) P_T/K} = \frac{1}{K-1} \geq t^* = \frac{1}{K-1}, \label{Equ:satisfied_constraint_one}
\end{equation}
which satisfies \eqref{Equ:min_SIR_constraint}. Furthermore, the total power constraint holds:
\begin{equation}
    \sum_{i=1}^{K} P_i = \sum_{i=1}^{K} \frac{P_T}{K} = P_T. \label{Equ:satisfied_constraint_two}
\end{equation}
Since the constraints are satisfied and the upper bound is achieved at equality, the uniform allocation is optimal, completing the proof.}

\ifCLASSOPTIONcaptionsoff
  \newpage
\fi

\bibliographystyle{IEEEtran}
\bibliography{references}

\begin{IEEEbiography}[{\includegraphics[width=1in,height=1.25in,clip,keepaspectratio]{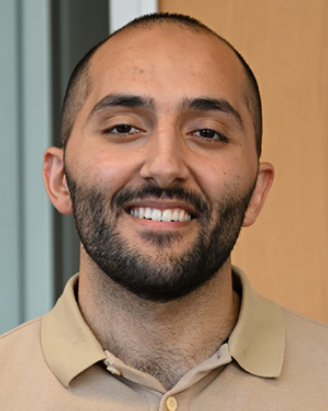}}]{Salah Abdeljabar}
(Graduate~Student~Member, IEEE) received the B.Sc. degree in electrical engineering from The University of Jordan, Amman, Jordan, in 2019, and the M.Sc. degree in electrical and computer engineering from the King Abdullah University of Science and Technology, Thuwal, Saudi Arabia, in 2023, where he is currently pursuing the Ph.D. degree. His research interests include long-range (LoRa) communication, Delay delay-tolerant networking (DTN), and optical wireless communications systems.
\end{IEEEbiography}

\begin{IEEEbiography}[{\includegraphics[width=1in,height=1.25in,clip,keepaspectratio]{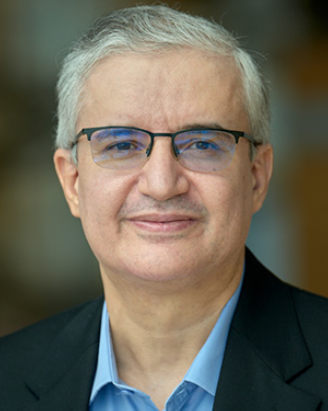}}]{Mohamed-Slim Alouini}
(Fellow, IEEE) was born in Tunis, Tunisia. He received the Ph.D. degree in electrical engineering from the California Institute of Technology, Pasadena, CA, USA, in 1998. He served as a Faculty Member with the University of Minnesota, Minneapolis, MN, USA, then with Texas A\&M University at Qatar, Doha, Qatar, before joining the King Abdullah University of Science and Technology, Thuwal, Makkah, Saudi Arabia, as a Professor of Electrical Engineering in 2009. His current research interests include the modeling, design, and performance analysis of wireless communication systems.
\end{IEEEbiography}

\end{document}